\documentclass[a4paper]{article}
\usepackage{geometry}
\usepackage{setspace}
\usepackage{float}
\usepackage{epstopdf}
\usepackage{standalone}
\usepackage{amsmath,amsfonts,amssymb}
\usepackage{color}

\usepackage{bm}
\usepackage{pdflscape}
\usepackage{authblk}
\usepackage{graphicx}
\usepackage[flushleft]{threeparttable}
\usepackage[comma,authoryear]{natbib}
\let\oldemptyset\emptyset

\definecolor{brown}{rgb}{0.59, 0.29, 0.0}
\usepackage{mathtools}

\newcommand{\interior}[1]{%
 {\kern0pt#1}^{\mathrm{o}}%
}
\usepackage [english]{babel}
\usepackage [autostyle, english = american]{csquotes}
\MakeOuterQuote{"}

\usepackage{algorithm,algpseudocode}
\usepackage{caption}
\usepackage[makeroom]{cancel}
\newcommand{\mbR}{{\mathbb R}}

\makeatletter
\newcommand*\bigcdot{\mathpalette\bigcdot@{.5}}
\newcommand*\bigcdot@[2]{\mathbin{\vcenter{\hbox{\scalebox{#2}{$\m@th#1\bullet$}}}}}
\makeatother

\title{\textbf{Bayesian Shrinkage for Functional Network Models, with Applications to Longitudinal Item Response Data}}

\author[1,2]{Jaewoo Park}
\author[1]{Yeseul Jeon}
\author[3]{Minsuk Shin}
\author[4]{Minjeong Jeon}
\author[1,2]{Ick Hoon Jin}
\affil[1]{Department of Statistics and Data Science, Yonsei University}
\affil[2]{Department of Applied Statistics, Yonsei University}
\affil[3]{Department of Statistics,  University of South Carolina}
\affil[4]{Graduate School of Education and Information Studies, University of California, Los Angeles}

\begin{document}

\maketitle

\begin{abstract}
Longitudinal item response data are common in social science, educational science, and psychology, among other disciplines. Studying the time-varying relationships between items is crucial for educational assessment or designing marketing strategies from survey questions. Although dynamic network models have been widely developed, we cannot apply them directly to item response data because there are multiple systems of nodes with various types of local interactions among items, resulting in multiplex network structures. We propose a new model to study these temporal interactions among items by embedding the functional parameters within the exponential random graph model framework. Inference on such models is difficult because the likelihood functions contain intractable normalizing constants. Furthermore, the number of functional parameters grows exponentially as the number of items increases. Variable selection for such models is not trivial because standard shrinkage approaches do not consider temporal trends in functional parameters. To overcome these challenges, we develop a novel Bayes approach by combining an auxiliary variable MCMC algorithm and a recently-developed functional shrinkage method. We apply our algorithm to survey and review data sets, illustrating that the proposed approach can avoid the evaluation of intractable normalizing constants as well as detect significant temporal interactions among items. Through a simulation study under different scenarios, we examine the performance of our algorithm. Our method is, to our knowledge, the first attempt to select functional variables for models with intractable normalizing constants.
\end{abstract}


\noindent%



\section{Introduction}

In many disciplines, including epidemiology, psychometrics, political science, and text mining, longitudinal item response model is widely used to analyze responses to test items collected over time. Examples include changes in the relationships among various attributes of hotel text review data \citep{han2016guests}, and longitudinal survey data for reporting the experiences of participants in psychology \citep{geschwind2011mindfulness}. This has practical implications; for instance, studying how the relationships among the sentiment keywords of hotel reviews change over time can be useful for designing marketing strategies. Network-based approaches are natural to describe the change in local interactions among items. However, it is not trivial to recover such interactions because the resulting network has multiple systems of nodes with various types of local interactions.

In this manuscript, we propose a new model to directly interpret the temporal interactions among items for longitudinal network data. We embed time-varying (or functional) interaction parameters within well-established exponential random graph model (ERGM) frameworks. Such models face several computational and inferential challenges: (1) the models include doubly-intractable normalizing constants, and (2) with an increasing number of items, the number of functional parameters grows exponentially. It is challenging to identify significant interaction parameters using standard variable selection approaches because those methods do not consider temporal trends in functional parameters. To address these challenges, we develop a novel Bayes approach based on the functional shrinkage method \citep{shin2019functional} combined with an auxiliary variable Markov chain Monte Carlo (MCMC) algorithm \citep{murray2006, liang2010double}. Our method can automatically detect strong temporal interactions among items, while avoiding the direct evaluation of the intractable normalizing constants included in the model. 

Exponential random graph models (ERGMs) \citep{robins2007introduction} are widely used to study the global structure of static networks. Bayesian approaches \citep{caimo2011bayesian,caimo2012bergm} are useful for ERGMs because they can alleviate the model degeneracy issue \citep{handcock2002statistical}; for some parameter region, ERGMs have high probability mass on either very dense or very sparse networks. Furthermore, the Bayesian framework is convenient for providing uncertainties from posterior samples. \cite{caimo2013bayesian} explores Bayesian model selection for ERGMs based on reversible jump MCMC. \cite{bouranis2017efficient} develops a correction method for pseudo-posterior samples to address computational issues for fitting large ERGMs. Recently, \cite{bouranis2018bayesian} proposes a fast Bayesian model selection in the ERGMs context by adjusting pseudolikelihood approximation. In the context of ERGMs, we focus here on a Bayesian approach to detect significant temporal interactions.

There is an extensive literature on temporal network modeling. These approaches can be broadly classified into two categories: (1) {\textit{exponential random graph models (ERGMs)}} \citep{hanneke2010discrete, krivitsky2014separable, lee2020varying} which describe the change in the topological structure of the networks, and (2) {\textit{latent space models (LSMs)}} \citep{sewell2015latent, friel2016interlocking, loyal2020bayesian} which embed change in transitive tendencies of nodes into low-dimensional latent space. In addition, dynamic network data may be classified as {\textit{unipartite}} or {\textit{bipartite}} networks. Unipartite dynamic networks have one set of nodes observed through $T$ time points, which result in $\mathbf{x} \in \mathbf{R}^{T \times n \times n}$ binary matrix. For all $l,j$, $x_{tlj}=1$ if the $l$-th node and $j$-th node are connected at time $t$; otherwise $x_{tlj}=0$. On the other hand, bipartite dynamic networks have two types of nodes; one set of nodes consists of the $n$ actors and the other sets have $p$ actors. These bipartite networks are observed over $T$ times resulting in $\mathbf{x} \in \mathbf{R}^{T \times n \times p}$ binary matrix; for all $l,j$, $x_{tlj}=1$ if the $l$-th actor in node type 1 connects to $j$-th actor in node type 2 at time $t$; otherwise $x_{tlj}=0$.

An item response data can be regarded as a bipartite network; we have $n$ respondents (node type 1) and $p$ items (node type 2). In this case, $x_{tlj}=1$ if respondent $l$ answers correctly (positively) to item $j$; otherwise $x_{tlj}=0$ at time $t$. It is challenging to apply existing dynamic network models to longitudinal item response data because most of them have been developed for unipartite networks. Furthermore, we have to consider various types of local interactions among items carefully. For instance, pairwise interactions between two items can be different from that of the other two items; therefore, assuming only one type of interaction between items (i.e. a single two-star statistics) is unrealistic. In this context, there have been several recent proposals for network psychometrics to study cross-sectional item response data sets \citep{nature14, jin2019doubly, park2019bayesian}. These provide an elegant approach to detect interactions among $p$ items from $n$ respondents. By extending these approaches, we focus on studying the temporal network structures among $p$ items for longitudinal item response data.

We propose functional inhomogeneous exponential random graph models (FI-ERGMs) that can detect significant temporal interactions between items. There are several inferential and computational challenges for FI-ERGMs. The likelihood functions involve intractable normalizing constants, which result in doubly-intractable distributions \citep{murray2006} in Bayesian analysis. Furthermore, with an increasing number of items ($p$), the number of model parameters grows at order $p^2$. Since these parameters are time-varying functional objects, the conventional variable selection method can not be directly applicable. Even in static networks, relatively few approaches have been developed for variable selection. Recently, \cite{nature14} imposes an $l_1$-penalty on Ising graphical models \citep{ravikumar2010high}. However, this method cannot quantify the uncertainty about the estimated interaction and is not robust to model misspecification. To address this, \cite{park2019bayesian} develops a Bayesian model selection approach for item response data. However, to our knowledge, no existing approaches provide a variable selection procedure for longitudinal networks. Furthermore, applying existing dynamic network models to longitudinal item response data is still limited. This motivates the development of new methods that allow shrinkage for time-varying local interactions. To address these challenges, we adopt two methods: (1) a double Metropolis-Hastings \citep{liang2010double}, an auxiliary variable MCMC method that cancels out the intractable normalizing constants in the acceptance probability, and (2) a functional horseshoe prior \citep{shin2019functional} that encourages shrinkage toward the zero function for weak signals. We show that our methods can recover the dependence structure of longitudinal networks as well as detect strong time-varying interactions. 

The rest of this manuscript is organized as follows. In Section 2, we describe existing dynamic network models and motivate a new model. Then we propose functional inhomogeneous exponential random graph models (FI-ERGMs) and discuss their computational and inferential challenges. In Section 3, we describe the functional horseshoe double Metropolis-Hastings (FHS-DMH) to make an inference for the FI-ERGMs and describe the implementation details. We provide a guideline to shrink the functional parameters, which can automatically detect the significant temporal interactions. In Section 4, we apply the FHS-DMH to three real data sets: Korea youth panel survey data, motivation to succeed survey data, hotel review data. In Section 5, we investigate the performance of the proposed method through simulation studies. We conclude with a discussion in Section 6. 

\section{Model}

\subsection{Dynamic Network Models}

\begin{table}[htbp]
\centering
\begin{tabular}{ccrrr}\hline
\multicolumn{2}{c}{} & Network types & Purpose of modeling \\
\hline
ERGM & TERGM & unipartite &  prevalence of ties \\
& STERGM  & unipartite & incidence and duration of ties\\
& VCERGM  & unipartite & time-varying topological features \\
\hline
LSM & DLSM  & unipartite  & latent positions  \\
& DLSM-B  & bipartite & latent positions \\
& HDP-LPCM  & unipartite & time-varying community structures \\
\hline
\end{tabular}
\caption{Comparison between dynamic network models. }
\label{t5}
\end{table}

In this section, we describe existing dynamic network models and point out the motivation for a new network model for analyzing longitudinal item response data. There have been a number of recent proposals for dynamic network modeling, summarized in Table \ref{t5}. 

\paragraph{Exponential Random Graph Models} 
Exponential random graph models (ERGMs) \citep{robins2007introduction} are widely used to study static networks. By extending ERGMs, dynamic models have been developed for unipartite networks. These models mainly focus on explaining the global structure with the network statistics $g(\mathbf{x})$. Here, we introduce three models as follows. 

\begin{enumerate}
\item TERGMs \citep{hanneke2010discrete} can account for temporal changes of networks by assuming a Markov-dependent structure on two time-adjacent networks. Let $\mathbf{x}_t$ be the $n \times n$ unipartite network observed at time $t$. TERGMs use transition statistics $g(\mathbf{x}_t, \mathbf{x}_{t-1})$ to describe the temporal change of the global structure of sequential networks. For instance, one can explain how the number of clusterings changes using triangle statistics; the positive model parameters indicate that it is more likely to have triangles in the observed network. Note that TERGMs can be generalized to a higher-order time dependence structure.

\item TERGMs focus on modeling the prevalence of the network at the current time point, such as the total number of connections (ties) at time $t$. However, it is also essential to study the incidence and duration of ties in dynamic networks. Furthermore, in many social science applications, factors that result in the incidence of ties and their duration are different. To make such interpretation, STERGMs \citep{krivitsky2014separable} introduce the formation and dissolution terms separately in the transition statistics. Compared to TERGMs, which use a single parameter to describe the cross-sectional properties of a network, STERGMs use formation and duration parameters to consider longitudinal properties of dynamic networks. 

\item 
Recently proposed varying-coefficient ERGM (VCERGM) \citep{lee2020varying} can parametrize the time-varying topological structure of networks in continuous time. Similar to TERGMs or STERGMs, VCERGMs use transition statistics to account for the change in the network structures. However, VCERGMs represent the corresponding model parameters as basis expansion, allowing to study the temporal heterogeneity of dynamic networks. Furthermore, VCERGMs can interpolate for unobserved networks because model parameters are smooth functions over continuous time. 
\end{enumerate}

\paragraph{Latent Space Models} 
Another common approach to study networks is the latent space model (LSM) \citep{hoff2002latent}. Such models embed network information into the low-dimensional Euclidean space so-called latent space. The closer two nodes in this latent space are, the more likely they are connected in the network. By modeling the evolution of latent positions over time, we can understand the change of local and global structures of dynamic networks. We introduce three recent approaches as follows. 

\begin{enumerate}
\item Dynamic latent space model (DLSM) \citep{sewell2015latent} allows each node to have a temporal trajectory in the Euclidean latent space. DLSM describes the formation of ties based on the distance between latent positions by modeling them as a hidden Markov process. The closer latent positions for two nodes are, the more likely they are connected in the observed dynamic network. Note that DLSM can be applicable to both undirected and directed networks by modeling sender and receiver in the formation of ties. 

\item While DLSM focuses on studying unipartite networks, \cite{friel2016interlocking} develops a dynamic latent space model for the bipartite networks, referred as DLSM-B. DLSM-B models the formation and duration of ties based on the latent distances from two different node types. \cite{friel2016interlocking} applied their model to dynamic bipartite networks consisting of $n$ companies (node type 1) and $p$ directors (node type 2), and capture the heterogeneity of dynamic bipartite networks.

\item Hierarchical Dirichlet process latent position clustering model (HDP-LPCM) \citep{loyal2020bayesian} can describe the time-varying community structures such as deletions, splits, or merges of groups in dynamic networks. Similar to other latent space-based models, HPD-LPCM assigns nodes' information into the latent space. In addition, HDP-LPCM can cluster the latent positions by incorporating the hierarchical Dirichlet process prior. From this, HDP can describe both the local and global structures in dynamic unipartite networks.
\end{enumerate}

\paragraph{Motivation} 
In item response theory, detecting significant interactions among items is crucial for designing or analyzing questionnaires. For instance, in clinical trial surveys, recovered interactions among symptoms can be used to diagnose patient groups for clinical intervention. Note that such interactions arise locally; interactions between two items can be different from that of the other two items. Therefore, a network model for item response data should allow such inhomogeneous dependence structures.

Although numerous dynamic network models are well established, direct application of these methods to longitudinal item response data set is challenging.  Most of the existing models have been developed for unipartite networks. Bipartite networks commonly arise in psychology and educational sciences. Item response data comprises complex network structures with two node types. For instance, we have survey data from $n$ respondents (node type 1) for $p$ items (node type 2) observed through $T$ time points. A notable exception is DLSM-B \citep{friel2016interlocking}, which can embed bipartite network information into latent space. Based on the latent positions of items, DLSM-B describes both local and global structures of networks. The closer latent positions are, the more likely items have positive interactions. 

DLSM-B can provide useful insights of these item interactions; however, it is not trivial to assess whether such interactions are significant or not: \textit{how} much should latent positions become closer to have statistically significant interactions? Our model can rule out weak interactions among items using the functional shrinkage \citep{shin2019functional}. Furthermore, our method can quantify temporal interactions through estimated model parameters, which is challenging for DLSM-B; we can only observe the change in interactions based on the relative distances between latent positions. In Section 4, we compare our model with DLSM-B to point out such differences. Our model belongs to ERGM class and has similarities to VCERGM \citep{lee2020varying} in that we use a functional representation of model parameters. However, we focus on modeling the change in local interactions among items rather than the evolution of the connectivity pattern of networks as in the standard ERGMs. Furthermore, FI-ERGM is proposed for bipartite networks from longitudinal item response data sets, while VCERGM is applicable to unipartite networks. We provide details about the difference between the standard ERGMs and I-ERGMs in Section 2.2.

\subsection{Functional Inhomogeneous Exponential Random Graph Models}

Let $\mathbf{x} \in \mbR^{T \times n \times p}$ denote  data with $n$ responses to $p$ binary items observed through $T$ time points. For all $l,j$, $x_{tlj}=1$ if the $l$-th individual responses $j$-th item positively (or correctly) at time $t$; otherwise $x_{tlj}=0$. To account for the pairwise temporal interactions among items, we propose the functional version of I-ERGMs \citep{FoSd86} as 

\begin{equation}
f(\mathbf{x} | \boldsymbol{\theta}) = \prod_{t=1}^{T} \frac{1}{\kappa(\boldsymbol{\theta}_{t \bigcdot })}\exp\left\{ \sum_{\forall j}^{p}\alpha_{tj}\sum_{l=1}^{n}x_{tlj} + \sum_{\forall j<k}^{p}\gamma_{tjk}\sum_{l=1}^{n}x_{tlj}x_{tlk}  \right\},~~~\bm{\theta}_{t \bigcdot}=\lbrace \lbrace \alpha_{tj} \rbrace_{\forall j}, \lbrace \gamma_{tjk} \rbrace_{\forall j< k} \rbrace.
\label{DIERGM}
\end{equation}
\noindent For time $t$, $\alpha_{tj}$ is an item easiness parameter, which acts as the intercept term for item $j$ and $\gamma_{tjk}$ is the pairwise interaction among $j,k$ items.  Consider the model parameter $\bm{\theta} \in \mbR^{T \times q}$ , where $q = p + p(p-1)/2$. We can define $\bm{\theta}_{t \bigcdot}$ and  $\bm{\theta}_{\bigcdot i}$ as denoting the $t$-th row of $\bm{\theta}$ and $i$-th column of $\bm{\theta}$, respectively. Then, $\bm{\theta}_{t \bigcdot} \in \mbR^{q}$ is the parameter for time $t$ and $\bm{\theta}_{\bigcdot i} \in \mbR^{T}$ is the $i$-th functional parameter over time. Our models assume that $\mathbf{x}_t$ (the observed binary response at time $t$) is only dependent on $\boldsymbol{\theta}_{t \bigcdot }$. In Section 3, we introduce functional priors to account for the temporal trends within each functional parameter $\bm{\theta}_{\bigcdot i}$.

The standard ERGMs are applicable to the single network, which assumes only one type of interaction between nodes. For instance, when we use edge and two-star statistics, standard ERGMs assume that these statistics have equal probabilities of being occurred across different nodes. Therefore, all the nodes' information is incorporated into a network statistic to explain the global structure of a single network (e.g., overall connectivity patterns). However, such an assumption is not realistic for item response data sets, which have multiplex network structures \citep{nature14,park2019bayesian}. Here, a network is represented as multiple systems of a set of nodes, and there can be various types of local interactions among nodes. Especially in item response data, we have two node types: $n$ respondents (node type 1) and $p$ items (node type 2). Furthermore, there are complex local interactions among items. For instance, pairwise interactions between $j,k$ items can be different from that of $j',k'$ items ($j \neq j', k \neq k'$). I-ERGMs are applicable to such multiplex networks by modeling $\gamma_{tjk}$ and $\gamma_{tj'k'}$ separately. From this, I-ERGMs allow the different probability of occurrences for network statistics. Therefore I-ERGMs are useful to explain the local behaviors of networks and are suitable for item response data.

Generally, the standard ERGMs suffer from model degeneracy and projectivity issues \citep{shalizi2013projectivity}. This phenomenon occurs when a single change of a dyad status significantly impacts the dyadic dependent statistics in the standard ERGMs. Here, highly impacted dyadic dependent statistics are dominant in  Monte Carlo simulations of networks, and it causes model degeneracy.
Similarly, the standard ERGM cannot guarantee the projectivity \citep{shalizi2013projectivity}. It is very rare to observe the entire network of interest; we can fit ERGMs to the observed sub-network (sample), and then infer the entire network (population) through the fitted model. Since a single change of a dyad status can make a huge influence on the other dyadic dependent statistics, observed dyadic dependent network statistics can vary drastically depending on whom is selected as a sample in the observed sub-network.

Compared to the standard ERGMs, FI-ERGM can avoid model degeneracy issue and guarantee the projectivity of the model. FI-ERGM analyzes temporal pairwise interaction networks (item-item networks) constructed from an item response ${\bf x}\in \mbR^{T \times n \times p}$. To construct an item-item network ${\bf A}_{t}\in \mbR^{p \times p}$ at time $t$, we use 
\[ {\bf A}_t = \bigg\{A_{tjk}\bigg\} = \bigg\{ \sum_{l=1}^n x_{tlj} x_{tlk} \bigg\}, \]
where $x_{tlj}$ and $x_{tlk}$ are responses from a respondent $l$ for item $j$ and $k$ at time $t$, respectively. In the case of $x_{tlk} = 1$,
\[\mbox{the change of } A_{tjk} = \left\{\begin{array}{rl} -1 & \mbox{if $x_{tlj}$ changes from 1 to 0} \\
1 & \mbox{if $x_{tlj}$ changes from 0 to 1,} \end{array}\right.\] 
If $x_{tlk} = 0$, there is no change in $A_{tjk}$. This implies the status change of item $j$ for respondent $l$ at time $t$ will make a very minimal influence on $A_{tjk}$ so that FI-ERGM does not suffer from model degeneracy. In a similar fashion, the observed network statistics do not change much depending on whom are included in samples; FI-ERGM can yield a consistent result.

One can generalize (1) by adding third-order interactions of the form $x_{tlj}x_{tlk}x_{tlm}$ or higher-order interactions with additional computational and statistical challenges. However, it is particularly difficult to add higher-order interactions to FI-ERGMs due to the highly correlated high-dimensional parameters. For instance, including third-order interactions to FI-ERGMs would result in $p+p(p-1)/2+p(p-1)(p-2)/6$ functional parameters. This requires a sufficient number of respondents ($n$) for accurate statistical inference; collecting such large longitudinal item response data is challenging in practice. Furthermore, such functional parameters are highly correlated. Let $\delta_{tjkm}$ be the third-order interaction parameters from $x_{tlj}x_{tlk}x_{tlm}$. If items $j,k,m$ have positive second-order interactions (i.e., $\gamma_{tjk}, \gamma_{tjm}, \gamma_{tkm} >0$), $\delta_{tjkm}$ should be positive. FI-ERGMs with higher-order interaction can suffer from the slow mixing of the Markov chain due to such highly correlated functional parameters. For this practical reason, recent proposals in psychometrics \citep{nature14, park2019bayesian} have adapted high-dimensional Ising models \citep{ravikumar2010high} to study pairwise interactions among items. Both approaches can accurately recover the $p \times p$ item-item network structures in educational assessment data. By extending these works, we focus on modeling second-order functional parameters to capture the temporal change in local pairwise interactions in the $p \times p$ item-item graph.

Inferences for FI-ERGMs are challenging because of the intractable normalizing constants $\kappa(\boldsymbol{\theta}_{t \bigcdot })$ included in \eqref{DIERGM}. At each time $t$, the calculation of $\kappa(\boldsymbol{\theta}_{t \bigcdot })$ requires summing the overall $2^{np}$ configurations of the binary response data, which is intractable even with the moderate sizes of $n$ and $p$. Although frequentist method \citep{nature14} for such models, it cannot provide the uncertainty of estimates and is not robust to model misspecification \citep{park2019bayesian}.

Another difficulty with FI-ERGMs is that the number of model parameters increases at an order of $p^2$ ($p$ represents the number of items). To rule out weak interaction parameters, \cite{park2019bayesian} develops a Bayesian variable selection method for I-ERGMs in static networks. However, compared with their problem, our case is more complex because of the functional parameters $\lbrace \bm{\theta}_{\bigcdot i} \rbrace^{q}_{i=1}$ that vary across each time point. To address this challenge, we propose a novel MCMC approach based on a shrinkage prior on function spaces. Our method can automatically detect functional parameters with a weak signal, while providing posterior samples from an intractable likelihood function. 

\section{Functional Horseshoe Double Metropolis-Hastings}

For FI-ERGMs, Bayesian approaches are useful to capture the dependence structure in temporal networks because we can easily incorporate shrinkage priors to rule out parameters with weak signals. In this section, we propose an MCMC algorithm for FI-ERGMs. We combine double Metropolis-Hastings (DMH) \citep{liang2010double} with the functional horseshoe prior \citep{shin2019functional} to address the computational and inferential challenges in FI-ERGMs. 

\subsection{Bayesian Hierarchical Models with the Functional Horseshoe Prior}

Since the number of model parameters ($q$) for FI-ERGMs increases exponentially, one might consider standard shrinkage priors \citep[cf.][]{George:93,carvalho2010horseshoe} to rule out weak signals. However, such methods do not account for the temporal trends in the functional parameters $\lbrace \bm{\theta}_{\bigcdot i} \rbrace^{q}_{i=1}$ in FI-ERGMs. Furthermore, it is not clear how to select the functional parameters through standard shrinkage priors because they can only induce sparsity on individual $\theta_{ti}$. Therefore, in this manuscript, we apply the functional horseshoe (FHS) prior \citep{shin2019functional}, which can impose shrinkage on the shape of functions. The FHS can encourage shrinkage toward any parametric class of functions. Here, we focus on shrinkage toward zero functions to detect strong signals, which allows us to perform a natural model selection in FI-ERGMs. \cite{shin2019functional} shows that the posterior constructed from the FHS prior is concentrated at a near-optimal min-max rate. 

Consider the $i$-th functional model parameter $\bm{\theta}_{\bigcdot i} \in \mbR^{T}$ in \eqref{DIERGM}. Then, the prior on the functional model parameter can be 
\begin{equation}
\bm{\theta}_{\bigcdot i} | \bm{\beta}_{i} \sigma^{2}_{i}  \sim N(\bm{\Phi \beta}_{i}, \sigma^{2}_{i}\mathbf{I}_{T} ),
\label{prior}
\end{equation}
\noindent where $\bm{\Phi}\in \mbR^{T\times k_n}$ is a matrix of prespecified basis functions and $\bm{\beta}_i \in \mbR^{k_n}$ is a vector of basis coefficients. $\sigma^{2}_{i}$ explains an error that cannot be captured by a mean trend $\bm{\Phi \beta}_{i}$. We assume that the nonparametric basis expansion can capture the temporally dependent trends within each functional model parameter $\bm{\theta}_{\bigcdot i}$. Here, we use the B-spline basis \citep{de1978practical}, but other basis functions can also be used. The FHS can shrink $\bm{\Phi \beta}_{i}$ toward the null function subspace spanned by a null regressor matrix $\bm{\Phi}_{0}$ with $d_0=\mbox{rank}(\bm{\Phi}_0)$. Then, we can define the FHS hyperpriors as

\begin{equation}
\begin{split}
\bm{\beta}_{i}|\sigma^{2}_{i},\tau_{i} & \propto (\sigma^{2}_{i}\tau^{2}_{i})^{-(k_{n}-d_{0})/2} \exp\Big \lbrace -\frac{1}{2\sigma^{2}_{i}\tau^{2}_{i}} \bm{\beta}^{'
}_{i} \bm{\Phi}'(\mathbf{I}-\mathbf{Q}_{0})\bm{\Phi\beta}_{i} \Big \rbrace, \\
\tau_{i} &\propto \frac{(\tau^{2}_{i})^{b-1/2}}{(1+\tau^{2}_{i})^{a+b}}1_{0,\infty}(\tau_{i}).
\end{split} 
\label{hyperprior}
\end{equation}
\noindent To impose shrinkage toward the zero function, we set the null function space as $\bm{\Phi}_{0}=\lbrace \oldemptyset \rbrace$. Then, $d_0=0$ and $\mathbf{I}-\mathbf{Q}_{0} = \mathbf{I}$, where $\mathbf{Q}_{0}=\bm{\Phi}_{0}(\bm{\Phi}_{0}^{'}\bm{\Phi}_{0})^{-1}\bm{\Phi}_{0}$. Following \cite{shin2019functional}, we set $a=1/2$ and $b=\exp\lbrace -k_n \log T/2 \rbrace$ to satisfy the near optimal nonparametric posterior contraction rate. We can summarize the hierarchical models for FI-ERGMs as  
    
\begin{equation}
 \pi(\boldsymbol{\theta},\bm{\beta},\bm{\tau},\bm{\sigma}^2 | \mathbf{x}) \propto f(\mathbf{x} | \boldsymbol{\theta}) \pi(\bm{\theta} | \bm{\beta},\bm{\sigma}^2)\pi(\bm{\beta}|\bm{\tau})\pi(\bm{\tau})\pi(\bm{\sigma}^2),
\label{posterior1}
\end{equation}
where
\begin{equation}
\begin{split}
f(\mathbf{x} | \boldsymbol{\theta}) &= \prod_{t=1}^{T} \frac{1}{\kappa(\boldsymbol{\theta}_{t \bigcdot})}\exp\left\{ \sum_{\forall j}^{p}\alpha_{tj}\sum_{l=1}^{n}x_{tlj} + \sum_{\forall j<k}^{p}\gamma_{tjk}\sum_{l=1}^{n}x_{tlj}x_{tlk}  \right\},\\
 \bm{\theta}_{\bigcdot i} | \bm{\beta}_{i} \sigma^{2}_{i} & \sim N(\bm{\Phi \beta}_{i}, \sigma^{2}_{i}\mathbf{I}_{T} ),\\
 \bm{\beta}_{i}|\sigma^{2}_{i},\tau_{i} & \propto (\sigma^{2}_{i}\tau^{2}_{i})^{-(k_{n}-d_{0})/2} \exp\Big \lbrace -\frac{1}{2\sigma^{2}_{i}\tau^{2}_{i}} \bm{\beta}^{'
}_{i} \bm{\Phi}'(\mathbf{I}-\mathbf{Q}_{0})\bm{\Phi\beta}_{i} \Big \rbrace, \\
\tau_{i} &\propto \frac{(\tau^{2}_{i})^{b-1/2}}{(1+\tau^{2}_{i})^{a+b}}1_{0,\infty}(\tau_{i}),\\
\sigma^{2}_{i} &\sim \mbox{IG}(1/100,1/100).\\
\end{split} 
\label{posterior2}
\end{equation}

\subsection{Markov chain Monte Carlo Implementation}

Our model \eqref{posterior2} includes intractable normalizing constants $\kappa(\bm{\theta}_{t \bigcdot})$, which pose inferential and computational challenges. The resulting posterior \eqref{posterior1} is called a doubly-intractable distribution having extra unknown normalizing terms $\kappa(\bm{\theta}_{t \bigcdot})$, which cannot be canceled out in standard MCMC approaches. Several Bayes methods have been developed for sampling from doubly intractable distributions. Just few of these include constructing Monte Carlo approximations for $\kappa(\bm{\theta}_{t \bigcdot})$ \citep[cf.][]{atchade2008bayesian,lyne2015russian, park2019function} and generating auxiliary variables to avoid the direct evaluation of $\kappa(\bm{\theta}_{t \bigcdot})$ \citep[cf.][]{murray2006,liang2010double}. Considering that constructing Monte Carlo approximations is unstable with an increasing number of parameters, auxiliary variable approaches may be appropriate for the problems considered in this manuscript. In particular, double Metropolis-Hastings (DMH) \citep{liang2010double} is the most practical for computationally expensive problems and the only feasible approach for high-dimensional parameter problems among current approaches (see \cite{park2018bayesian} for comparisons between algorithms). Therefore, in what follows, we incorporate FHS priors with DMH to impose functional shrinkage as well as estimate the model parameters for FI-ERGMs.    

Consider the model parameter $\bm{\theta}$, and the hyperparameters for the FHS priors $(\bm{\beta},\bm{\tau},\bm{\sigma}^2)$. We update the parameters sequentially through the Gibbs sampler. Let the parameters at the $m$-th iteration of the Markov chain be 
\begin{equation}
\Big(\boldsymbol{\theta}^{(m)}, \boldsymbol{\beta}^{(m)}, \bm{\tau}^{(m)}, \bm{\sigma}^{2(m)}\Big) = \Big(\bm{\theta}^{(m)}_{\bigcdot 1}, \cdots \bm{\theta}^{(m)}_{\bigcdot q}, \bm{\beta}^{(m)}_{1}, \cdots, \bm{\beta}^{(m)}_{q},\tau^{(m)}_{1},\cdots,\tau^{(m)}_{q},\sigma^{2(m)}_{1},\cdots,\sigma^{2(m)}_{q}\Big).
\label{parameters}
\end{equation}
\noindent For $i=1,\cdots,q$ we use the arbitrary initial values with $\bm{\theta}_{\bigcdot i}^{(0)} \sim \mbox{Uniform}(-5,5), \bm{\beta}_i \sim N(0,\mathbf{I}_{k_{n}}), \tau_i \sim N(0,1/\sqrt{20}), \sigma^{2}_{i} = 1$. We can update the parameters successively. The $i$-th model parameter for time $t$ can be updated from 
\begin{equation}
\theta_{ti}^{(m+1)} \sim  \pi\Big(\theta_{ti}^{(m)}| \mathbf{x}_t, \bm{\theta}_{t(i)}^{(m)}, \bm{\beta}_{i}^{(m)},\sigma^{2(m)}_{i}\Big) \propto f\Big(\mathbf{x}_t|\theta_{ti}^{(m)},\bm{\theta}_{t(i)}^{(m)}\Big)\pi\Big(\theta_{ti}|\bm{\beta}_{i}^{(m)},\sigma^{2(m)}_{i}\Big),
\label{thetasample}
\end{equation}
where $\boldsymbol{\theta}_{t(i)}^{(m)} = \Big(\theta_{t1}^{(m+1)} \cdots, \theta_{t i-1}^{(m+1)}, \theta_{t i+1}^{(m)}, \cdots, \theta_{t q}^{(m)}\Big)$. Since $f\Big(\mathbf{x}_t|\theta_{ti}^{(m)},\bm{\theta}_{t(i)}^{(m)}\Big)$ includes intractable $\kappa(\bm{\theta}_{t})$, we use a double Metropolis-Hastings (DMH) algorithm \citep{liang2010double} to update $\theta_{ti}^{(m+1)}$. This is a nested MCMC algorithm; a Metropolis-Hastings sampler is implemented within another Metropolis-Hastings sampler. At each iteration of the MCMC (outer MCMC) $\theta_{ti}^{'}$ is proposed from the proposal $q\Big(\cdot|\theta_{ti}^{(m)}\Big)$. For the given $\theta_{ti}^{'}$, DMH simulates an auxiliary variable $\mathbf{y}_{t}$ from the probability model $f\Big(\mathbf{x}_{t} \mid  \theta'_{ti}, \boldsymbol{\theta}_{t(i)}^{(m)}\Big)$ through the standard Metropolis-Hastings sampler (inner sampler). For each iteration of the inner sampler, $(j,k)$ pairs from $\mathbf{x}_{t}$ are randomly chosen; $x_{tjk}$ is set to $0$ or $1$ based on the full conditional probabilities of the networks. See \cite{hunter2008ergm} for more details. Theoretically, we can simulate an exact auxiliary variable as the inner sampler length approaches infinity; of course the length of the inner sampler should be finite in practice. Following \cite{liang2010double}, we use the inner sampler length as $2n$, where $n$ is the number of respondents for each $\theta_{ti}$ update. Considering that we have $q$ model parameters with $T$ times points, this choice results in a $2nqT$ number of MH updates to generate auxiliary variables. In Section 5, we study the performance of the algorithm with different lengths of the inner sampler. We observe that $2n$ can generate a reasonably accurate auxiliary variable. Then, the resulting acceptance probability for updating $\theta_{ti}^{(m+1)}$ is 
\begin{equation}
\alpha = \min\left\lbrace \frac{f\Big(\mathbf{x}_t \mid \theta_{ti}', \bm{\theta}_{t(i)}^{(m)}\Big) f\Big(\mathbf{y}_t \mid \theta_{ti}^{(m)},\bm{\theta}_{t(i)}^{(m)}\Big)\pi\Big(\theta_{ti}'|\bm{\beta}^{(m)}_{i}, \sigma^{2(m)}_{i}\Big)}{ f\Big(\mathbf{x}_t \mid \theta_{ti}^{(m)},\bm{\theta}_{t(i)}^{(m)}\Big)f\Big(\mathbf{y}_t|\theta_{ti}', \bm{\theta}_{t(i)}^{(m)}\Big) \pi\Big(\theta^{(m)}_{ti} \mid \bm{\beta}^{(m)}_{i}, \sigma^{2(m)}_{i}\Big)}
, 1 \right\rbrace.
\label{dmhacc}
\end{equation}
We note that \eqref{dmhacc} does not include intractable normalizing constant $\kappa(\bm{\theta}_t)$. The main idea of this approach is to cancel out $\kappa(\bm{\theta}_t)$ in the acceptance probability with a clever choice of the auxiliary variable. The more the simulated $\mathbf{y}_t$ is close to the observed $\mathbf{x}_t$, the more likely the proposed $\theta_{ti}^{'}$ will be accepted. We repeat this procedure for $i=1,\cdots,q$ and $t=1,\cdots,T$.

Then, for $i=1,\cdots,q$ $\tau^{(m+1)}_{i}$ can be obtained from \begin{equation}
\tau^{(m+1)}_{i}  \sim \pi\Big(\bm{\beta}^{(m)}_i \mid \sigma^{2(m)}_{i}, \tau^{(m)}_{i}\Big)\pi\Big(\tau^{(m)}_{i}\Big).
\label{tausamp}
\end{equation}
We transform $\eta_{i}^{(m)} = \tau_{i}^{-2(m)}$ to use a slicer sampler for better mixing as well as computational efficiency. The remaining parameters $\sigma^{2(m+1)}_{i}$ and $\bm{\beta}^{(m+1)}_{i}$ can be updated from 
\begin{equation}
\begin{split}
 \sigma^{2(m+1)}_{i} & \sim \pi\Big(\bm{\theta}^{(m+1)}_{\bigcdot i} \mid \bm{\beta}^{(m)}_{i}, \sigma^{2(m)}_{i}\Big) \pi\Big(\bm{\beta}^{(m)}_{i} \mid \sigma^{2(m)}_{i}, \tau^{(m+1)}_{i}\Big)\pi\Big(\sigma^{2(m)}_{i}\Big)\\
 \bm{\beta}^{(m+1)}_{i} & \sim \pi\Big(\bm{\theta}^{(m+1)}_{\bigcdot i} \mid \bm{\beta}^{(m)}_{i}, \sigma^{2(m+1)}_{i}\Big) \pi\Big(\bm{\beta}^{(m)}_{i} \mid \sigma^{2(m+1)}_{i}, \tau^{(m+1)}_{i}\Big),
\end{split}
\label{betasigmasamp}    
\end{equation}
where the conditional distributions are an inverse gamma distribution and a normal distribution, respectively. The conditional distributions for all parameters are described in the supplementary material. The FHS-DMH algorithm is summarized in Algorithm~\ref{FHSDMHalg}. 

\begin{algorithm}[htbp]
\caption{A Functional Horseshoe Double Metropolis-Hastings (FHS-DHM) Algorithm}\label{FHSDMHalg}
\begin{algorithmic}
\normalsize

\State \textbf{Given $\boldsymbol{\theta}^{(m)}, \boldsymbol{\beta}^{(m)}, \bm{\tau}^{(m)}, \bm{\sigma}^{2(m)}$ update $\boldsymbol{\theta}^{(m+1)}, \boldsymbol{\beta}^{(m+1)}, \bm{\tau}^{(m+1)}, \bm{\sigma}^{2(m+1)}$} for all $i=1,\cdots, q$ and $t=1,\cdots, T$.\\

\State {\it{Step 1.}} Propose $\theta'_{ti} \sim q\Big(\cdot|\theta^{(m)}_{ti}\Big)$.\\

\State {\it{Step 2.}} Generate an auxiliary variable from the probability model using the $2n$-number of Metropolis-Hastings updates:\\
$\mathbf{y}_{t} \sim f\Big(\mathbf{x}_{t} \mid  \theta'_{ti}, \boldsymbol{\theta}_{t(i)}^{(m)}\Big)$, where $\boldsymbol{\theta}_{t(i)}^{(m)} = \Big(\theta_{t1}^{(m+1)} \cdots, \theta_{t i-1}^{(m+1)}, \theta_{t i+1}^{(m)}, \cdots, \theta_{t q}^{(m)}\Big)$.\\

\State {\it{Step 3.}} Accept $\theta^{(m+1)}_{ti}=\theta'_{ti}$ with probability
$$\alpha = \min\left\lbrace \frac{  f\Big(\mathbf{x}_t|\theta_{ti}',\bm{\theta}_{t(i)}^{(m)}\Big)f\Big(\mathbf{y}_t|\theta_{ti}^{(m)},\bm{\theta}_{t(i)}^{(m)}\Big)\pi\Big(\theta_{ti}'|\bm{\beta}^{(m)}_{i}, \sigma^{2(m)}_{i}\Big)
}{  f\Big(\mathbf{x}_t|\theta_{ti}^{(m)},\bm{\theta}_{t(i)}^{(m)}\Big)f\Big(\mathbf{y}_t|\theta_{ti}',\bm{\theta}_{t(i)}^{(m)}\Big)\pi\Big(\theta^{(m)}_{ti}|\bm{\beta}^{(m)}_{i}, \sigma^{2(m)}_{i}\Big)}
, 1 \right\rbrace$$ else reject (set $\theta^{(m+1)}_{ti}=\theta^{(m)}_{ti}$).\\

\State {\it{Step 4.}} Update $\tau^{(m+1)}_{i}$ using a slice sampler :\\\\

\begin{enumerate}
\item $u \sim \mbox{Uniform} \bigg[ 0, \Big( \frac{1}{1+\eta_{i}^{(m)}} \Big)^{a+b} \bigg]$, where $\eta_{i}^{(m)} = \tau_{i}^{-2(m)}$.

\item  $\eta_{i}^{(m+1)} \sim \mbox{Uniform} \bigg[0,u^{a + k_{n}/2 - d_{0}/2 -1 } e^{ -\frac{1}{2\sigma^{2(m)}_{i}} \bm{\beta}^{(m)'
}_{i} \bm{\Phi}'(\mathbf{I}-\mathbf{Q}_{0})\bm{\Phi}\bm{\beta}^{(m)}_{i} u} \bigg]$. 

\item Take $\tau_{i}^{(m+1)} = \frac{1}{\sqrt{\eta_{i}^{(m+1)}}}$.
\end{enumerate}\\\\

\State {\it{Step 5.}} Update $\sigma^{2(m+1)}_{i}$ from the conditional distribution:\\
$$\sigma^{2(m+1)}_{i} \sim \mbox{IG}\Big(T/2+k_{n}/2 +1/100, \bm{\beta}^{(m)'
}_{i} \bm{\Phi}'\bm{\Phi}\bm{\beta}^{(m)}_{i}/2 + \bm{\beta}^{(m)'
}_{i} \bm{\Phi}'(\mathbf{I}-\mathbf{Q}_{0})\bm{\Phi}\bm{\beta}^{(m)}_{i}/2\tau^{2(m+1)}_{i}+1/100\Big)$$\\

\State {\it{Step 6.}} Update $\bm{\beta}^{(m+1)}_{i}$ from the conditional distribution:\\
$$\bm{\beta}^{(m+1)}_{i} \sim \mbox{N}\bigg( \Big[\bm{\Phi}'\bm{\Phi} + \mathbf{I}(1/\tau^{2(m+1)}_{i})\Big]^{-1} \bm{\Phi}'\bm{\theta}^{(m+1)}_{\bigcdot i}, \sigma^{2(m+1)}_{i}\Big[\bm{\Phi}'\bm{\Phi} + \mathbf{I}(1/\tau^{2(m+1)}_{i})\Big]^{-1}\bigg)$$

\end{algorithmic}
\end{algorithm}

With each iteration of the MCMC, our algorithm generates an auxiliary variable for $qT$ times to update $\lbrace \theta_{ti} \rbrace$ (see Step 2 of Algorithm~\ref{FHSDMHalg}). Since we use the inner sampler length as $n$ (identical to the number of respondents), the computational complexity of the brute force implementation is $\mathcal{O}(np^2 T)$, where $p$ is the number of items and $T$ is the number of observed time points. However, in \eqref{posterior2}, we assume that $\mathbf{x}_t$ is only dependent on $\boldsymbol{\theta}_{t \bigcdot }$. Therefore, we can use parallel computing to generate $\mathbf{x}_{t}$ independently of the given $\bm{\theta}_{t \bigcdot}$. Then the computational complexity of our method is $\mathcal{O}(np^2 T/c)$, where $c < T$ is the number of available processors. The parallel computation is implemented through the {\tt{OpenMp}} library in {\tt{C++}}. FHS-DMH can automatically shrink the parameters $\lbrace \bm{\theta}_{\bigcdot i} \rbrace_{i=1}^{q}$ having weak signals to zero functions as well as generate posterior samples from the complex hierarchical models in \eqref{posterior2}.

\subsection{Shrinkage Procedure for Functional Parameters in FI-ERGMs}

For FI-ERGMs, we have the functional parameters $\lbrace \bm{\theta}_{\bigcdot i} \rbrace_{i=1}^{q}$, where $q=p+p(p-1)/2$ increases exponentially with the number of items ($p$). Among the $q$ functional parameters, the first $p$ of them represent the easiness of the corresponding items across $T$ time points. These can be regarded as the intercept terms in standard regression models. The remaining $p(p-1)/2$ number of parameters represent the pairwise interaction among the items across $T$ time points. Since not every item has statistically significant interactions, it is important to identify the important interactions only by shrinking the others to zero. 

Consider the reparameterization $\omega_i = 1/(1+\tau_{i}^{2})$, where $\omega_i$ can be interpreted as the weight that the posterior mean for function places on the null function subspace \citep{shin2019functional}. Therefore, a larger $\omega_i$ indicates higher weights on the zero function for $\bm{\theta}_{\bigcdot i}$. According to \cite{shin2019functional}[Theorem 3.2], the posterior distribution of $\omega_i$ converges toward 1 when the true shape of $\bm{\theta}_{\bigcdot i}$ is the zero function. On the contrary, if the true $\bm{\theta}_{\bigcdot i} \neq \mathbf{0}$, the posterior distribution of $\omega_i$ contracts toward 0. Following \cite{shin2019functional}, we set the threshold for $\omega_i$ as 0.5. If the posterior mean of $\omega_i$ is greater than 0.5, we diagnose  $\bm{\theta}_{\bigcdot i} = \mathbf{0} \in \mbR^{T}$; otherwise $\bm{\theta}_{\bigcdot i} \neq \mathbf{0} \in \mbR^{T}$

\section{Applications}

Here, we illustrate the application of our method to three real data examples: (1) Korea youth panel survey data, (2) motivation to succeed survey data, and (3) hotel review data. We observe that FHS-DMH can shrink weak interactions toward the zero function as well as recover the dependence structure of the longitudinal network well. For the MCMC implementation, we use an independent normal proposal. The convergence of MCMC methods has been checked by the Monte Carlo standard errors \citep{jones2006fixed, flegal2008markov}. 

To validate our method, we compare the summary statistics between the observed network and fitted network from the posterior predictive distribution \citep{gelman2013bayesian}. For observed data $\mathbf{x}$, we use the following summary statistics  
\[
T(\mathbf{x}) = \left\lbrace \Big\lbrace \sum_{l=1}^{n}x_{tlj} \Big\rbrace_{\forall j}, \Big\lbrace \sum_{l=1}^{n}x_{tlj}x_{tlk} \Big\rbrace_{\forall j<k} \right\rbrace_{\forall t}.
\]
Here, $T(\mathbf{x}) \in R^{q \times T}$ because we have $q=p+p(p-1)/2$ parameters for $T$ time points. For a given posterior sample from FHS-DMH, we simulate binary response data $\mathbf{y}$. Then we obtain summary statistics $T(\mathbf{y})$. If these synthetic summary statistics $T(\mathbf{y})$ resemble the observed summary statistics $T(\mathbf{x})$ well, then our FHS-DMH posterior sample can be regarded as a reasonable approximation of the true posterior distribution. To implement this, we obtain 1,000 thinned posterior samples from FHS-DMH. We simulate $\mathbf{y}_1,\cdots,\mathbf{y}_{1,000}$ from the 1,000 posterior samples. Then, we calculate the sample mean of the summary statistics $\frac{1}{n}\sum_{i=1}^{n}T(\mathbf{y}_i)$. 

In addition, we assess our fitted models in terms of higher-order degree statistics. From the posterior predictive distribution, for each time $t$ we calculate the $p \times p$ item-item graph for each respondent $l$. A $j,k$ element of the item-item graph becomes 1, if a respondent $l$ gives a correct (positive) responses to both items $j,k$ at time $t$ (i.e., $x_{tlj} x_{tlk}=1$). Then we calculate the degree statistics (order $m$) in the item-item graph obtained from $n$ respondents and take the average of the degree statistics over $n$ respondents. Then we compare the sample mean of the simulated degree statistics with the observed degree statistics as before. We provide model validations using degree statistics of order $m=0,\cdots,p-1$ for $t=1,\cdots,T$, which results in $pT$ number of degree statistics. 

\subsection{Korea Youth Panel Survey Data}

\paragraph{Data Description} 
The first example came from the Korea Youth Panel Survey \citep{lee2010korea} that tracked a nationally representative sample of second-year middle school students for six consecutive years from 2003 to 2008. Following \cite{jeon2012profile}, we analyzed the subset of the data that excludes about 2\% of the students who changed their school membership during their middle school and/or high school years. 

In this application, we used the five years' data (2004--2008) on the 30 items that measure how the students think about themselves (i.e., self-image). The base year (2003)'s data were dropped because six items were not included in the base year. Each of the 30 item was measured on a 5-point Likert-type scale with the response options, ``Strongly disagree'', ``Disagree'', ``Neither agree nor disagree'', ``Agree'', and ``Strongly agree''. Example items include  I sometimes think I am a useless person, I sometimes think I am a bad person, I sometimes feel like I am a failure, and  I think I am a trouble make. A full set of items are provided in the supplementary material. To measure a positive self-image, the response categories of the negative items were reversed, and then all response categories were dichotomized at point 3 ($\leq$ 3 and $>$ 3). The proportion of male students was 49.9\%. At each time point, the data include binary responses from about $n=3,000$ individuals to the $p=30$ items; this results in $p+p(p-1)/2=465$ functional parameters in FI-ERGMs.

\begin{figure}[htbp]
\begin{center}
\includegraphics[width=0.95\textwidth]{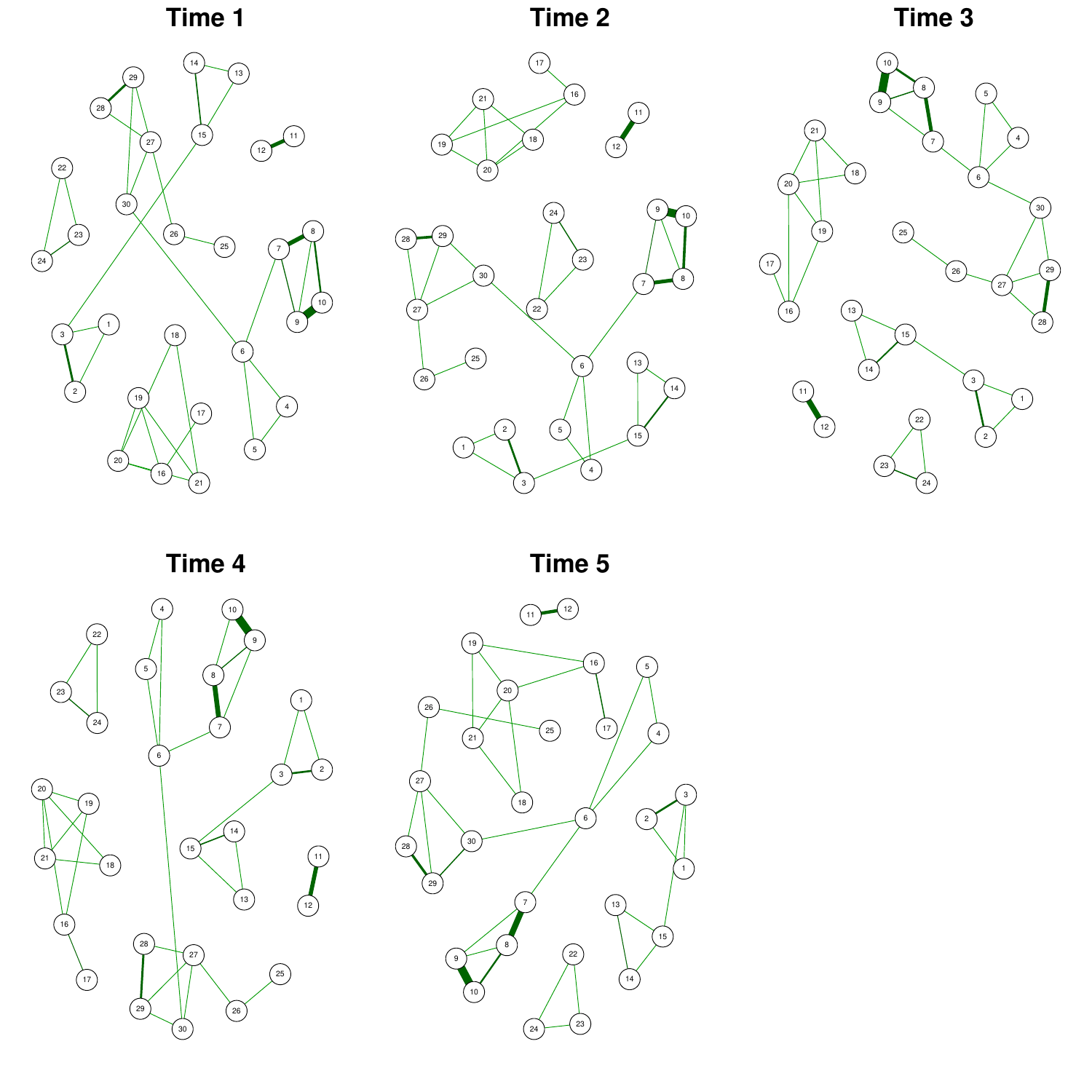}
\end{center}
\caption[]{Estimated networks for the Korea youth panel survey data set. Green lines indicate positive relations. The width of the lines indicate the connection strength between the relevant items - thicker lines indicate stronger interaction between items.}
\label{selfimagenetwork}
\end{figure}

\begin{table}[tt]
\centering
\begin{tabular}{crrrrrr} \hline
Positive  & Time 1 & Time 2 & Time 3 & Time 4 & Time 5 \\ 
\hline
$\gamma_{t,9,10}$  &  7.068 & 6.343 & 5.348 & 6.736 & 7.820\\
$\gamma_{t,7,8}$  & 4.559 & 3.771 & 3.438 & 4.754 & 6.176\\
$\gamma_{t,11,12}$  & 4.175 & 4.512 & 4.264 & 4.350 &4.241\\
$\gamma_{t,28,29}$ & 3.242 & 3.149 & 3.349 & 3.162 & 3.683\\
$\gamma_{t,2,3}$ & 3.133 & 2.771 & 2.734 & 3.047 & 3.263\\
$\gamma_{t,8,10}$ & 2.914 & 3.156 & 2.820 & 1.740 & 3.116\\
$\gamma_{t,14,15}$ & 2.594 & 2.501 & 2.536 & 2.626 & 2.324\\
$\gamma_{t,23,24}$ & 2.359 & 2.223 & 2.403 & 2.451 & 2.342\\
$\gamma_{t,16,17}$ & 2.175 & 2.003 & 2.165 & 2.323 & 2.646\\
$\gamma_{t,13,14}$ & 2.128 & 2.014 & 2.151 & 2.260 & 2.428\\
\hline
\end{tabular}
\caption{Top 10 largest nonzero interaction parameters among items in the Korea youth panel survey data. The order is based on the summation of the estimated interaction parameters across all time (i.e., $\sum_{\forall t} \gamma_{tjk}$). Estimates are obtained from posterior mean of 20,000 MCMC samples from FHS-DMH; the Monte Carlo standard errors are at 0.02.}
\label{selfimagetable} 
\end{table}

\begin{figure}[htbp]
\begin{center}
\includegraphics[width=1.0\textwidth]{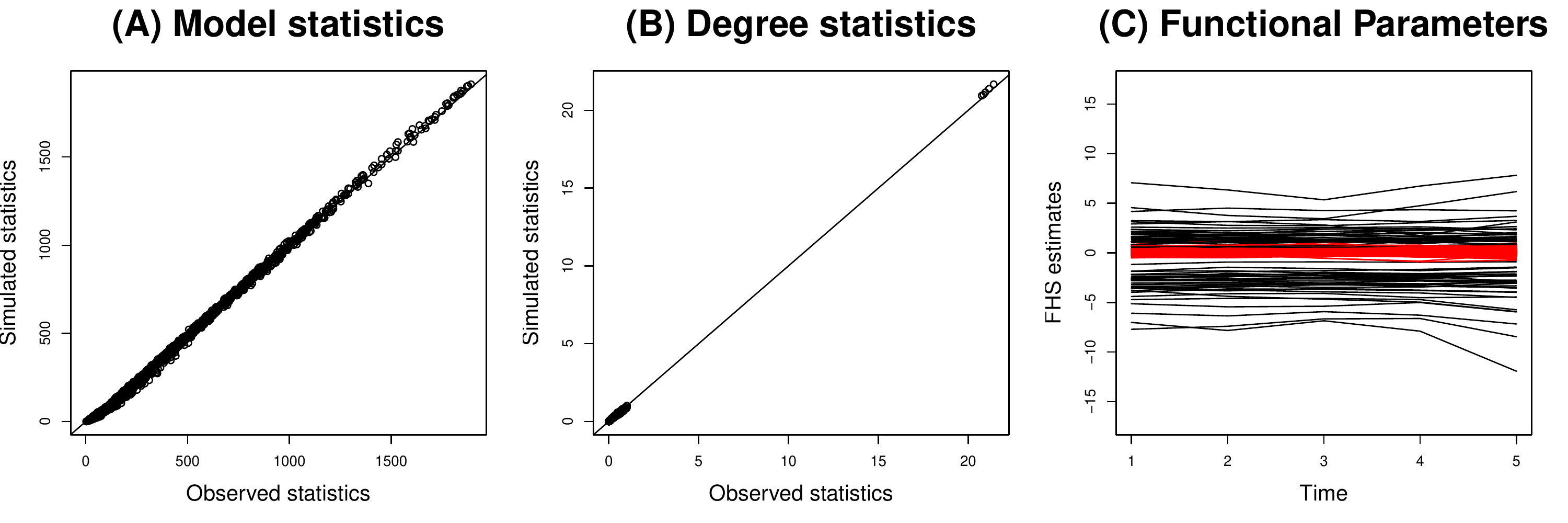}
\end{center}
\caption[]{The left panel (A) compares the observed and mean of the simulated model summary statistics. The middle panel (B) compares the observed and mean of the simulated degree statistics. The summary statistics are simulated 1,000 times for the given FHS-DMH estimates. The right panel (C) shows the shrinkage effect for the functional parameters. The red lines indicate shrinkage. Among the 465 functional parameters, 399 of them are diagnosed as the zero functions.}
\label{selfimagestats}
\end{figure}

\paragraph{Analysis Results}
Among the 465 functional parameters, FHS-DMH shrinks 399 of them to zero functions (Figure~\ref{selfimagestats} (C)). Our method takes about 28 hours. Figure~\ref{selfimagenetwork} and Table~\ref{selfimagetable} describe the estimated dependence structures and their nonzero interactions. We observe that every connection of items shows positive relationships, and the dependence structure is overall consistent over time. We provide some descriptive explanations based on the top 10 largest positive interactions in terms of the posterior mean of the parameters 
$\gamma_{t,j,k}$ shown in Table~\ref{selfimagetable}.

The strongest positive interaction occurs between items (9) (“Other people think I am a trouble maker”) and (10) (”Other people think I am a juvenile delinquent”), which 
makes sense given that  both items are about how other people think and `trouble maker' and `juvenile delinquent' are about similar (negative) image. The second strongest positive interaction
is shown between items (7) (“I think I am a trouble maker”) and (8) (“I think I am a juvenile delinquent”); the interaction between these items becomes stronger as time goes. 
These items are similar in content to the first pair of items (9) and (10) with the strongest interaction; the only difference is that this second pair is about how they think about themselves.  
The third strongest positive interaction appears between items (11) (“If I do something bad, people will blame me”)
and (12) (“If I do something bad, I'll be humiliated by other people”). 
This indicates that the teenagers who respond positively to items (11) and (12) are concerned about other people's negative  opinion about them. Therefore, they are less likely to cause troubles which could damage to their self-image. 
A full list of items of the self-image data set can be found in the supplement. In Figure~\ref{selfimagestats} (A) and (B), the mean of the observed statistics and simulated statistics follow a straight line, indicating that our model fits well.

\begin{figure}[htbp]
\begin{center}
\includegraphics[width = 1.0\textwidth]{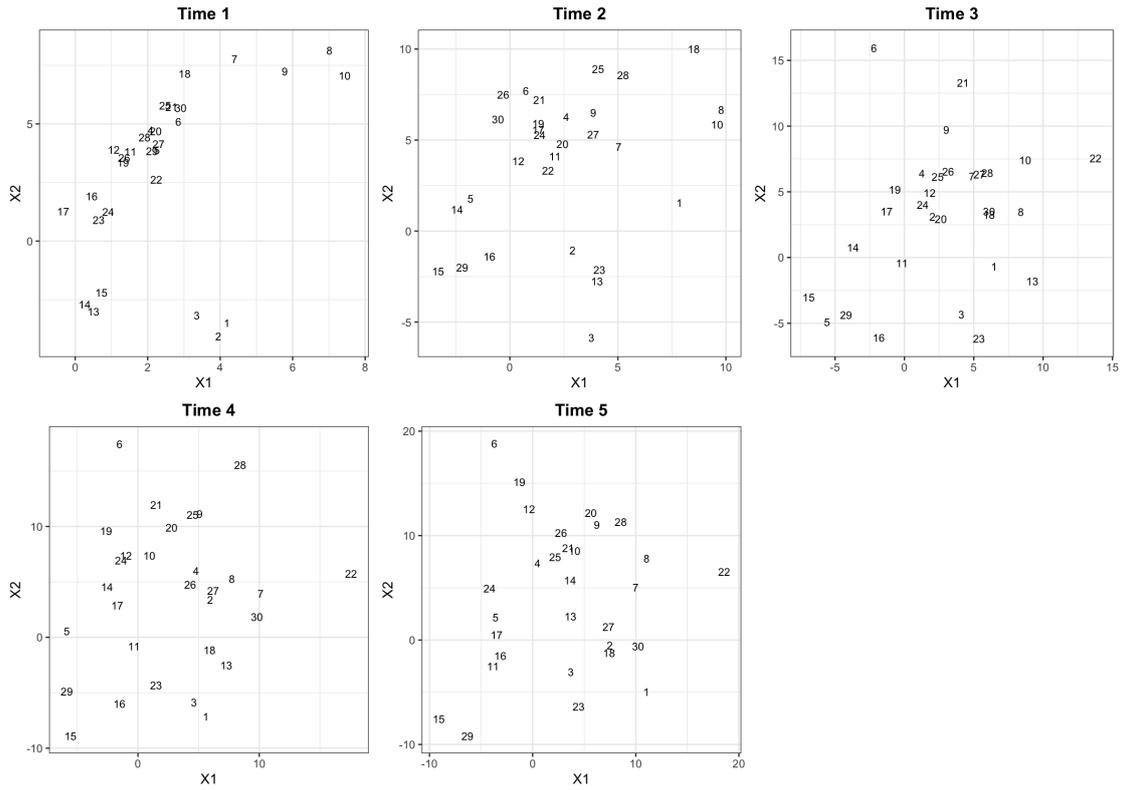}
\end{center}
\caption[]{Estimated latent positions for the Korea youth panel survey data set. The closer two latent positions of items are, the more likely they have positive relations. }
\label{selfimagedlsm}
\end{figure}

\paragraph{Comparison with DLSM-B}

Figure \ref{selfimagedlsm} represents the estimated latent positions from DLSM-B. Estimates are obtained from the posterior mean of 20,000 MCMC samples, which takes about 8 minutes. 
For comparison, we selected the two pairs that showed top two largest interactions with the proposed method (Table~\ref{selfimagetable}). 
The latent positions between the  item pair (9) and (10) as well as the pair (7) and (8)  are relatively close to each other within the pair across years,  
implying positive temporal interactions between the items within each pair. This finding suggests some consistency in the results between the two methods. 
To capture temporal changes in local interactions between items with DLSM-B, one needs to compare and identify changes in the relative distances between all item pairs in the latent space. These comparisons are somewhat cumbersome and decisions on close and distant distances may be seen somewhat subjective. In addition, with a large number of items, it can be challenging and inconvenient to determine meaningful changes over time in the the interactions (distances) between all item pairs  (Figure~\ref{selfimagedlsm}). On the other hand, FHS-DMH, our proposed approach, directly quantifies temporal interactions through estimated $\gamma_{t,j,k}$, and therefore, detects statistically significant interactions at the expense of computing time (Figure~\ref{selfimagenetwork}, Table~\ref{selfimagetable}).

\subsection{Motivation to Succeed Data}
\paragraph{Data Description} 
The data for the second application came from the Pathways to Desistance study \citep{mulvey2004theory}, which is a multi-site longitudinal study that follows 1,354 serious juvenile offenders from adolescence to young adulthood in Philadelphia and Phoenix. Participants completed baseline interviews in November 2000 and follow-up interviews at 6, 12, 18, 24, 30, 36, 48, 60, 72, and 84 months post-baseline (first follow up interview completed in May 2001; last follow up interview completed in March 2010). The aims of the study are to identify initial patterns of how serious adolescent offenders stop antisocial activity; describe the role of the social context and developmental changes in promoting these positive changes; and compare the effects of sanctions and interventions in promoting these changes. 


From this large study, we used the motivation to succeed scale \citep{eccles1998motivation}, which includes six items that measure the respondent's assessment of the opportunities available in his/her neighborhood regarding schooling and work. An additional two items were included that measure the adolescent's perceptions of how far he/she would like to go in school and how far he/she think they will go in school. The eight test items are as follows: 
(1) In my neighborhood, it is easy for a young person to get a good job; 
(2) Most of my friends will graduate from high school;  
(3) In my neighborhood, it is hard to make money without doing something illegal; 
(4) College is too expensive for most people in my neighborhood;  
(5) We have fewer opportunities to succeed than kids from other neighborhoods;  
(6) Our chances of getting ahead/being successful are not very good; 
(7) How far would you like to go in school?  
(8) How far do you think you will go in school?

The responses to the first six questions are on a five-point Likert-scale (``Strongly disagree'', ``Disagree'', ``Neither agree nor disagree'', ``Agree'', and ``Strongly agree''). The responses to the last two questions are ``Drop out before graduation'', ``Graduate from HS'', ``Go to business, tech school or jr college'', ``Graduate from college'', and ``Go to graduate or professional school''. We dichotomized the responses to the first six items by assigning 1 to  ``Agree'', and ``Strongly Agree'' responses and 0 otherwise. For the last two items, we assigned 1 to ``Graduate from college'' and ``Go to graduate or professional school'' and 0 otherwise. After removing cases with missing responses, 740 respondents remained. 
In summary, at each time point, the data include $n=740$ respondents for the $p=8$ items, which results in 36  functional parameters in FI-ERGMs.

\begin{figure}[htbp]
\begin{center}
\includegraphics[width=1.1\textwidth]{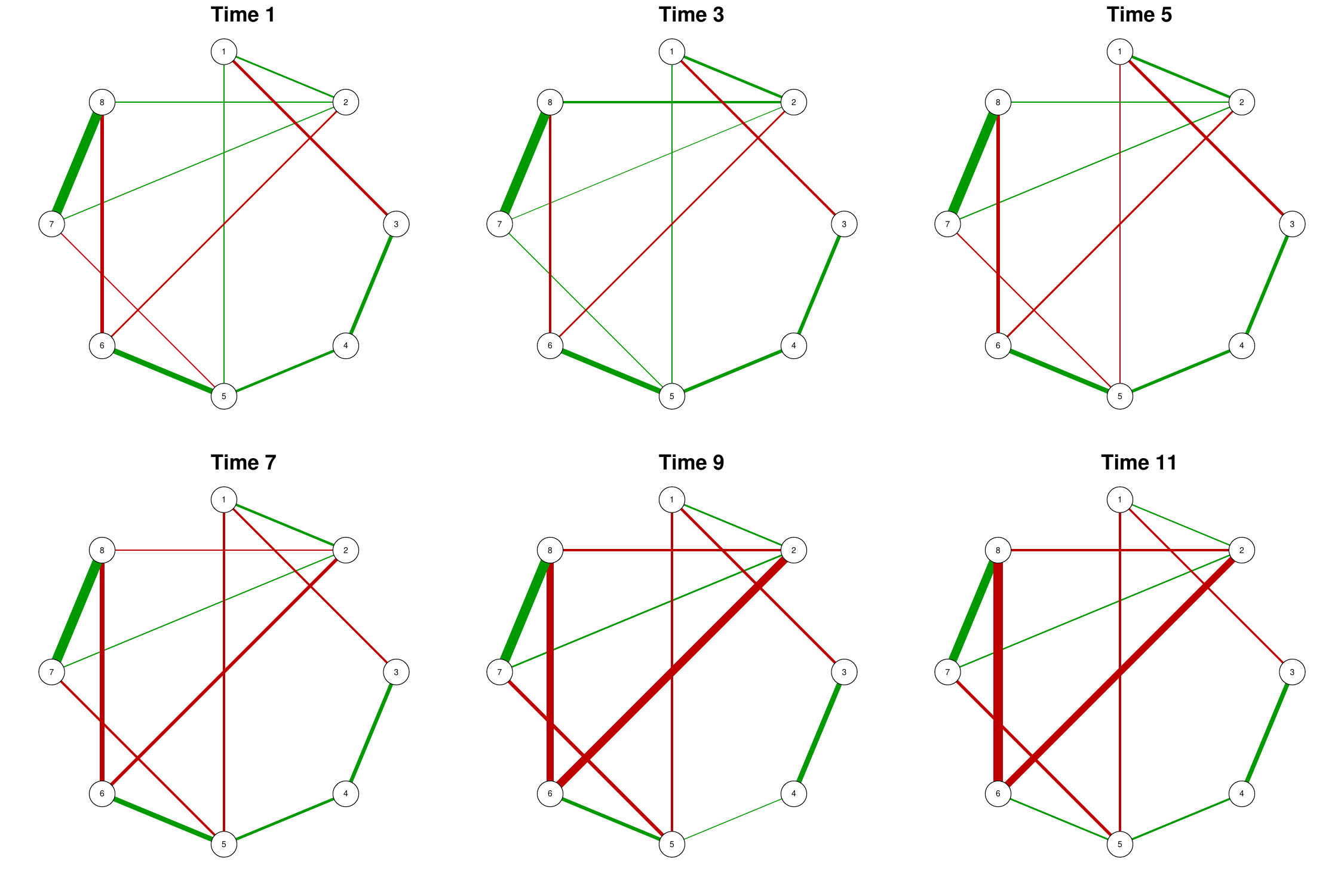}
\end{center}
\caption[]{Estimated networks for the motivation to succeed data set. The green lines indicate positive relations and the red lines represent negative relations. The width of the lines indicates the connection strength between the relevant nodes; thicker lines indicate stronger interactions between items.}
\label{motivatenetwork}
\end{figure}

\begin{table}[tt]
\centering
\begin{tabular}{crrrrrr}\hline
Interactions  & Time 1 & Time 3 & Time 5 & Time 7 & Time 9 & Time 11 \\ \hline
$\gamma_{t,1,2}$ & 0.470 & 1.069 & 1.047 & 0.961 & 0.610 & 0.453\\
$\gamma_{t,1,3}$ & -0.880 & -0.927 & -1.164 & -0.781 & -1.176 & -0.910\\
$\gamma_{t,3,4}$ & 1.353  & 1.595  & 1.518  & 1.681  & 2.716 &  2.762\\
$\gamma_{t,1,5}$ & 0.005  & 0.071 & -0.517 & -1.007 & -1.292 & -1.688\\
$\gamma_{t,4,5}$ & 0.869  & 1.479 & 1.252  & 1.108 & 0.113 & 0.966\\
$\gamma_{t,2,6}$ & -0.410 & -0.560 & -0.605 & -1.440 & -4.414 & -4.562\\
$\gamma_{t,5,6}$ & 2.128  & 2.584 &  2.145  & 2.387 & 1.684  & 0.765\\
$\gamma_{t,2,7}$ & 0.143 &  0.034  & 0.263 &  0.254 &  0.586 &  0.542 \\
$\gamma_{t,5,7}$ & -0.187 &  0.136 & -0.303 & -0.859 & -1.726 & -1.714 \\
$\gamma_{t,2,8}$ & 0.196 &  0.716 &  0.537 & -0.009 & -0.961 & -1.070 \\
$\gamma_{t,6,8}$ & -1.544 & -0.841 & -1.505 & -2.031 & -3.998 & -6.812 \\
$\gamma_{t,7,8}$ & 4.163 & 5.078 & 4.846  & 5.084 & 5.534  & 5.864 \\ \hline
\end{tabular}
\caption{Estimated nonzero interaction parameters among the items in the motivation to succeed data. The estimates are obtained from the posterior mean of 30,000 MCMC samples from FHS-DMH; Monte Carlo standard errors are at 0.04.}
\label{motivatetable} 
\end{table}

\begin{figure}[htbp]
\begin{center}
\includegraphics[width=1.0\textwidth]{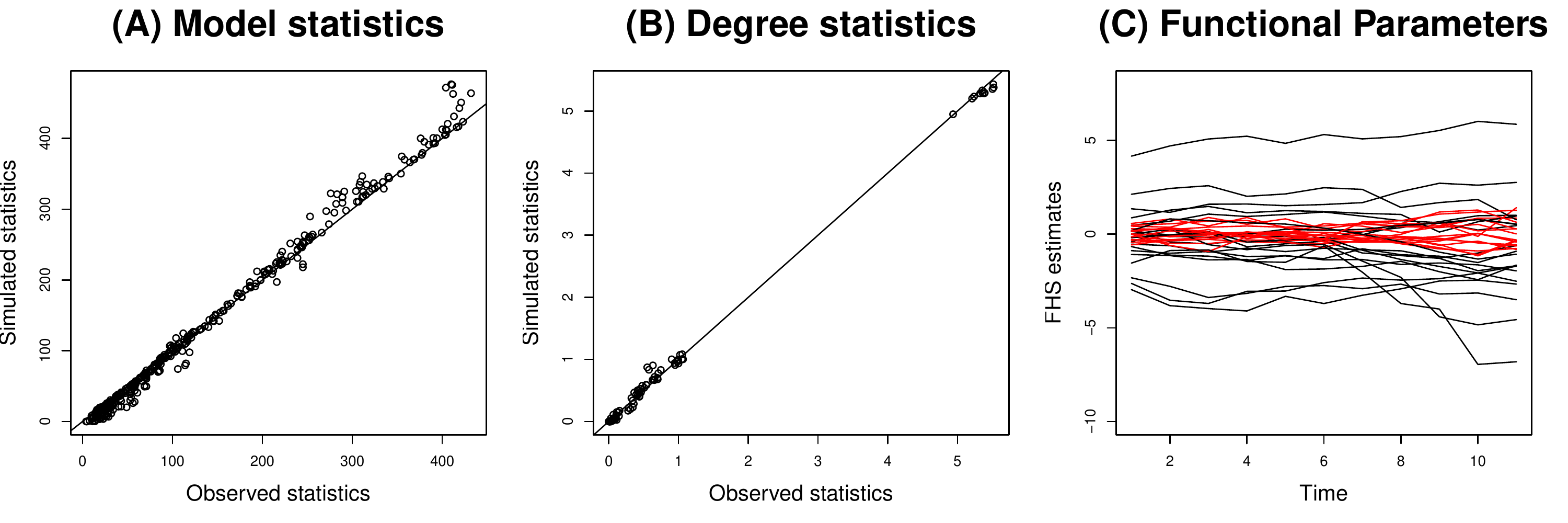}
\end{center}
\caption[]{The left panel (A) compares the observed and mean of the simulated model summary statistics. The middle panel (B) compares the observed and mean of the simulated degree statistics. The summary statistics are simulated 1,000 times for the given FHS-DMH estimates. The right panel (C) shows the shrinkage effect for the functional parameters. The red lines indicate shrinkage. Among the 36 functional parameters, 17 of them are diagnosed as the zero functions.}
\label{motivatestats}
\end{figure}

\paragraph{Analysis Results}
Figure~\ref{motivatestats} (B) shows that among the 36 functional parameters, FHS-DMH shrinks 17 of them to zero functions. Our method takes about an hour. Figure~\ref{motivatenetwork} and Table~\ref{motivatetable} show the estimated network structures and their estimated nonzero interaction parameters. We observe several important negative and positive connections among the items. Such relationships are generally consistent over time, whereas their strengths vary to some degree. In particular, item (1) and item (3) are negatively connected. This indicates that students who believe that it is easy for young people to get a job in their neighborhood tend to disagree that doing illegal things is necessary to earn money. On the contrary,  item (3) shows a positive relationship with item (4). Students who deem it necessary to engage in illegal activities to earn money also think that college tuition is too expensive. From these findings, we can conclude that poverty is an important factor behind students’ antisocial activities.

In addition, item (6), which represents the chance of success, has negative relationships with items (8) and (2), but a positive relationship with item (5). Such connections show that students who anticipate them having little chance of success are less likely to think that they will opt for higher education. Instead, they think their neighborhoods, including themselves, have fewer opportunities to be successful. These relationships convey that students who think they have insufficient opportunities than others are less likely to think that they will go college or graduate high school; rather, they think their neighborhood has fewer opportunities for them to be successful.

In particular, the strength of the interaction between item (6) and item (2) becomes stronger over time. It can be inferred from this result that earning a high school diploma or higher degree brings students closer to success. Therefore, education is an important factor in leading students to success. Additionally, items (7) and (8) have the strongest relationship; this indicates that a desire and willingness to go to college are closely related. This suggests that the desire to go to college motivates students to overcome the obstacles presented by their living situation, should those obstacles exist, and pursue their dream of higher education.

As in the previous example, we simulate the summary statistics from 1,000 thinned posterior samples obtained from FHS-DMH. Figure~\ref{motivatestats} shows that the mean of the 1,000 simulated summary statistics aligns with the observed summary statistics, indicating that our model fits the observed data well. 
\begin{figure}[htbp]
\begin{center}
\includegraphics[width = 1.0\textwidth]{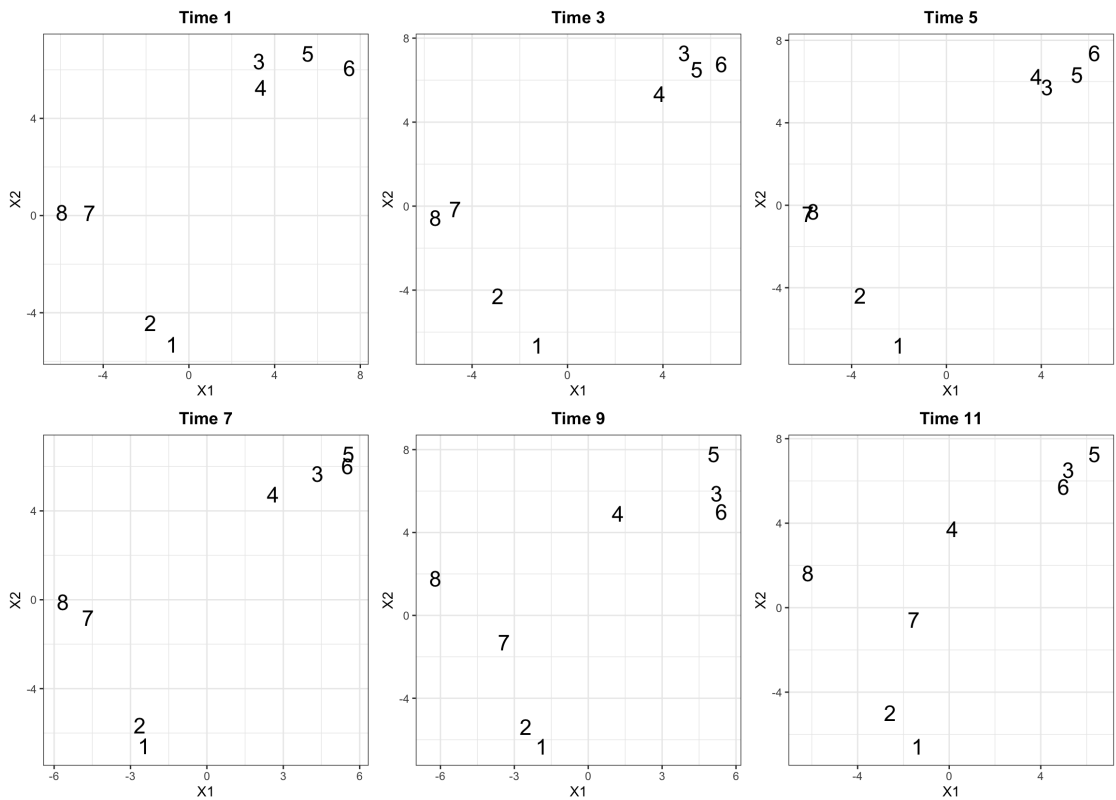}
\end{center}
\caption[]{Estimated latent positions for the motivation to succeed data set. The closer two latent positions of items are, the more likely they have positive relations. }
\label{motivedlsm-b}
\end{figure}

\paragraph{Comparison with DLSM-B} 
Figure \ref{motivedlsm-b} illustrates the estimated latent positions of items from DLSM-B. Estimates are obtained from the posterior mean of 10,000 MCMC samples, which takes about 6 minutes. There are three clusters of latent positions: items (1)-(2), items (3)-(6), and items (7)-(8). The closer latent positions are, the more likely they have positive interactions. These results are consistent with the estimated network from FHS-DMH (Figure~\ref{motivatenetwork}). Note that for DLSM-B estimates, it is difficult to distinguish between the zero and negative interactions. Since the latent positions of items (1), (8) are located far away from item (6), one can interpret that they have no interactions. On the other hand, FHS-DMH can quantify negative interactions (item (6) and item (8)) and zero interactions (item (1) and item (8)) separately (Table~\ref{motivatetable}, Figure~\ref{motivatenetwork}).

\subsection{Hotel Review Data}
\paragraph{Data Description}

\begin{table}[tt]
\centering
\begin{tabular}{cll}
  \hline
Item  & Aspect-based Sentiments & Sentiment Keywords \\
  \hline
1 & Price - satisfaction &  reasonable price, inexpensive price\\
2 & Price - dissatisfaction & very expensive, terrible price, ridiculous price \\
3 & Room - satisfaction & amazing room, luxury room, quiet room\\
4 & Room - dissatisfaction & dry room, dirty room, old room\\
5 & Subsidiary facilities - satisfaction & clean lounge, comfortable lounge\\
6 & Subsidiary facilities - dissatisfaction & dirty floor, old lounge\\
7 & Food - satisfaction & delicious buffet, nearby restaurant, clean cafeteria\\
8 & Food - dissatisfaction & dirty restaurant, tasteless cafeteria\\
9 & Interior design - satisfaction & nice place, luxurious interior\\
10 & Interior design - dissatisfaction & poor lighting, no sunlight\\
11 & Service - satisfaction & great service, prompt service\\
12 & Service - dissatisfaction & abysmal service, satisfactory service, surly service\\
13 & Bed - satisfaction & get a good night’s sleep\\
14 & Bed - dissatisfaction & terrible bed, uncomfortable bed\\
\hline
\end{tabular}
\caption{Aspect-based sentiments and their sentiment keywords for the hotel review data.}
\label{keywordsummary} 
\end{table}

Many reviewers express their opinions by writing reviews on websites, and their satisfaction toward hotels is summarized with sentiment keywords. Sentiment keywords represent emotional expressions in hotel review data; the positive and negative opinions in the review data correspond to specific keywords. We use text mining to construct aspect-based sentiments, which are composed of keywords with similar aspects. Here, we study the temporal networks among aspect-based sentiments about hotel reviews from reservation websites. We collect 231,862 reviews of 423 hotels in South Korea from 2018 to 2019 and exclude advertisements and spam reviews. To obtain the sentiment keywords from the data, we use the natural language process through the Daumsoft Text Mining Engine Version 2. With these keywords, we construct 14 aspect-based sentiments such as prices, rooms, subsidiary facilities, food, interior design, service, and bed. Table~\ref{keywordsummary} summarizes the aspects-based sentiments and their keywords. The data set contains binary review information from $n=423$ hotels over the seven time points from the first quarter of 2018  to the third quarter of 2019. The binary value indicates whether an individual hotel has a review containing aspect-based sentiments (1 existence of sentiments in the review and 0 otherwise). The $p=14$ aspect-based sentiments result in FI-ERGMs with 105 functional parameters.

\begin{figure}[htbp]
\begin{center}
\includegraphics[width=1.05\textwidth]{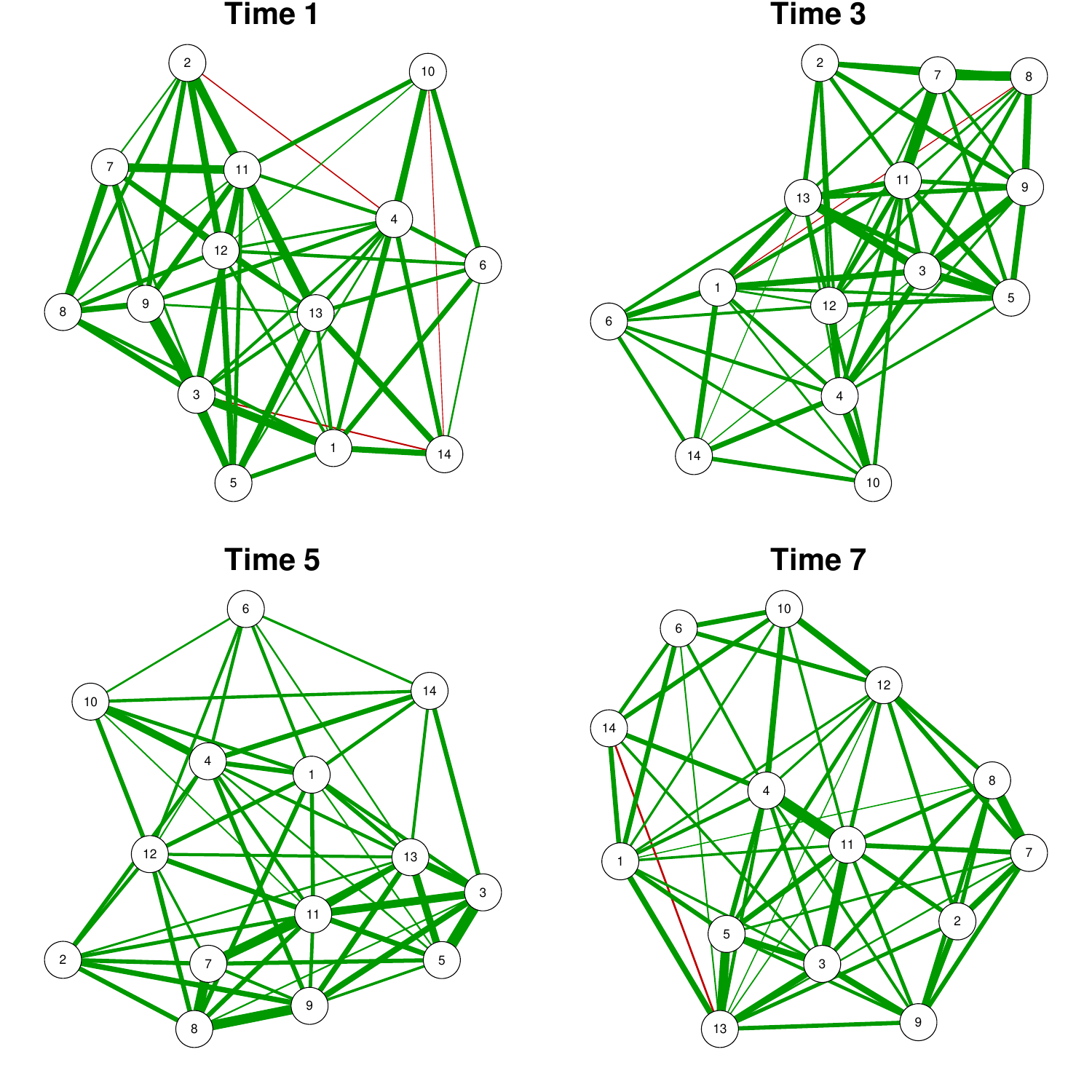}
\end{center}
\caption[]{Estimated networks for a hotel keyword data set. Green lines indicate positive relations and red lines represent for negative relations. The width of the lines indicate the connection strength between the relevant items - thicker lines indicate stronger interaction between keywords.}
\label{hotelnetwork}
\end{figure}

\begin{table}[tt]
\centering
\begin{tabular}{crrrrrrr} \hline
Interactions  & Time 1 & Time 2 & Time 3 & Time 4 & Time 5 & Time 6 & Time 7\\ \hline
$\gamma_{t,7,8}$  & 1.415 & 1.510 & 1.480 & 1.942 & 1.636 & 1.860 & 1.631\\
$\gamma_{t,7,11}$ & 1.543 & 1.271 & 1.792 & 1.461 & 1.487 & 1.036 & 0.775 \\
$\gamma_{t,3,5}$  & 1.293 & 1.217 & 1.144 & 1.151 & 1.891 & 1.564 & 1.037\\
$\gamma_{t,3,11}$ & 1.266 & 0.695 & 0.644 & 1.163 & 1.286 & 1.906 & 1.554\\
$\gamma_{t,5,13}$ & 1.259 & 0.847 & 0.796 & 1.069 & 1.188 & 1.606 & 1.603\\
$\gamma_{t,3,9}$  & 2.093 & 1.220 & 1.297 & 0.804 & 0.963 & 0.844 & 1.055\\
$\gamma_{t,8,9}$  & 1.036 & 1.004 & 1.141 & 1.378 & 1.629 & 1.080 & 0.717\\
$\gamma_{t,3,13}$ & 0.512 & 1.447 & 1.306 & 1.363 & 1.169 & 0.852 & 1.062\\
$\gamma_{t,4,10}$ & 1.257 & 1.092 & 1.167 & 0.963 & 1.224 & 0.816 & 0.981\\
$\gamma_{t,2,9}$  & 0.877 & 0.899 & 0.727 & 0.913 & 0.835 & 0.699 & 0.976\\
\hline
\end{tabular}
\caption{Top 10 largest nonzero interaction parameters among items in the hotel keyword data. The order is based on the summation of the estimated interaction parameters across all time (i.e., $\sum_{\forall t} \gamma_{tjk}$). Estimates are obtained from posterior mean of 50,000 MCMC samples from FHS-DMH; the Monte Carlo standard errors are at 0.03.}
\label{hoteltable2} 
\end{table}

\paragraph{Analysis Results}

\begin{figure}[htbp]
\begin{center}
\includegraphics[width=1\textwidth]{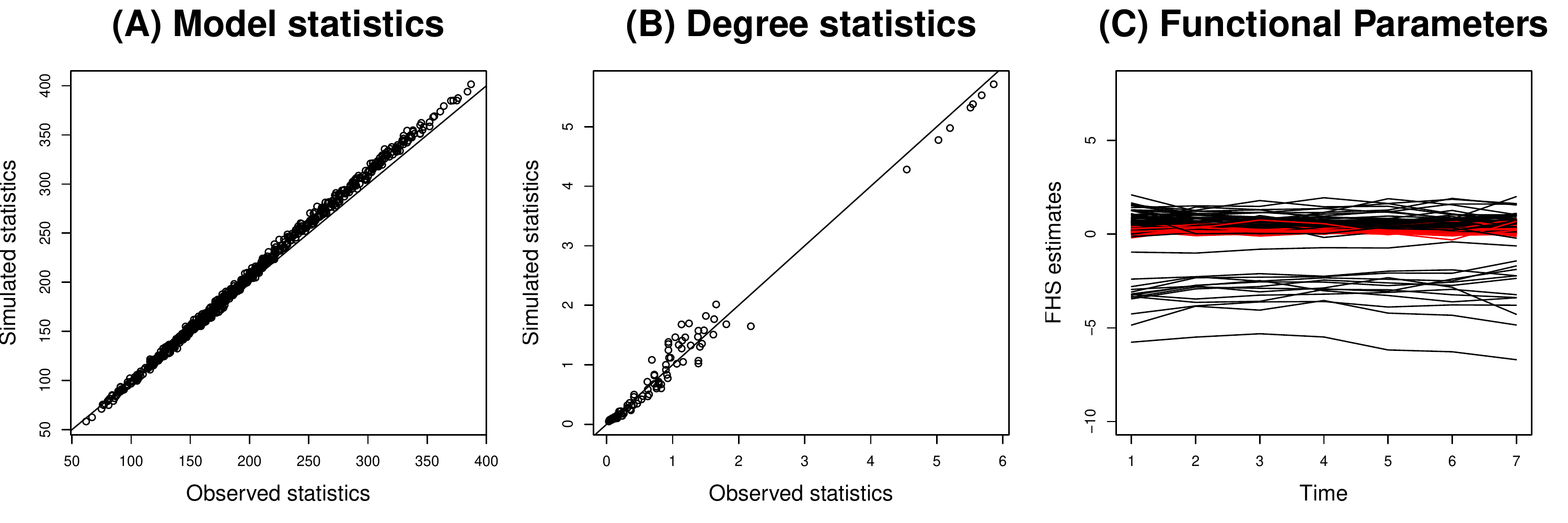}
\end{center}
\caption[]{The left panel (A) compares the observed and mean of the simulated model summary statistics. The middle panel (B) compares the observed and mean of the simulated degree statistics. The summary statistics are simulated 1,000 times for the given FHS-DMH estimates. The right panel (C) shows the shrinkage effect for the functional parameters. The red lines indicate shrinkage. Among the 105 functional parameters, 33 of them are diagnosed as the zero functions.}
\label{hotelstats}
\end{figure}

\paragraph{\textit{Analysis of Dyadic Relationships}}
Figure~\ref{hotelstats} (B) shows that among the 105 functional parameters, FHS-DMH shrinks 33 of them to zero functions. Figure~\ref{hotelnetwork} and Table~\ref{hoteltable2} describe the estimated network structures and their nonzero interaction parameters. FHS-DMH takes about 3 hours. In Figure~\ref{hotelnetwork}, we used the layout to the method of Fruchterman and Reingold \citep{fruchterman1991}, which exploits analogies between the relational structure in graphs, to visualize the node cluster structure efficiently. We provide a circular layout for the hotel review data in the supplementary material. We observe several meaningful patterns. First, item (8) (food dissatisfaction) and item (9) (interior design satisfaction) are connected positively at each time point. This indicates that although reviewers may be satisfied with the interior and exterior design of the hotel (including the restaurant), they are not necessarily satisfied with the taste of the food. On the contrary, satisfaction with the food (item (7)) is linked to satisfaction with the service (item (11)). Since this also has a strong positive relationship at all times, it can be seen as related to satisfaction with the food and service. However, Table~\ref{hoteltable2} shows that food satisfaction (item (7)) and food dissatisfaction (item (8)) has a strong positive relationship at all times, even though they have the opposite meanings. The items connecting favorable and unfavorable sentiments tend to display positive relationships because of the characteristics of the review data (i.e., reviewers mention negative and positive sentiment keywords on different aspects simultaneously in reviews). Therefore, the input matrix of the review data according to favorable and unfavorable aspect-based sentiments is analyzed with co-occurrence information. When this occurs, our model labels the interaction between satisfaction and dissatisfaction as positive. From this, we can infer that reviewers tended to say both good things and bad things about their food. Note that there is an exception; we observe a negative relationship between item (13) (bed satisfaction) and (14) (bed dissatisfaction) at Time 7. From this, we can conclude that once reviewers gratify their bed conditions, most reviewers less likely to respond the unfavorable reviews on their bed conditions.

\paragraph{\textit{Analysis of Negative Relationships}}
At Time 1, there is a negative relationship between room satisfaction (item (3)) and bed dissatisfaction (item (14)). We can infer that there are some favorable reviews about room condition but discontent with the bed quality for reasons such as a hard bed, dirty bed, etc. Moreover, based on the negative relationship between interior dissatisfaction (item (10)) and bed dissatisfaction (item (14)), some reviewers distinct unfavorable feedback of room interior from bed conditions. We can infer that even though customers dislike the interior of a room, such as poor lighting, they may satisfy the bed conditions or vise versa. In addition, we observe a negative relationship between item (2) (price dissatisfaction) and item (4) (room dissatisfaction) at Time 1. This implies that even if the price of a room is high, many reviewers content with room conditions. However, these negative connections turn into positive relationships over time.



\paragraph{\textit{Analysis of Triangle Relationships}}

There is a consistent triangle (cyclic) relationship between favorable aspect-based items over time. At Time 1, the connection between room satisfaction (item (3)) and interior design satisfaction (item (9)) is the strongest and their interaction parameter becomes smaller over
the seven time points. Furthermore, both those items are connected to satisfaction with the service (item (11)); hence the three items have a triangle relationship that is consistent at each time point. In other words, many reviews contain a favorable impression of the room, interior design, and service. Furthermore, there is a strong positive relationship between satisfaction with the food (item (7)) and satisfaction with the service (item (11)) at Time 3. In addition, these two items display a strong positive relationship with room satisfaction (item (3)) at all times. Given that items (3) and (11) are connected to item (5) (subsidiary facilities satisfaction), they form another triangle relationship. We can combine this result to conclude that hotel users have favorable opinions about their room, the service, the interior design, the food, and the subsidiary facilities. Considering that such triangle relationships are composed of items (3) and (11), we can infer that satisfaction with the service and with the room conditions are major factors in getting favorable reviews. 

As shown in Figure~\ref{hotelnetwork} and Table~\ref{hoteltable2}, the triangle relationships between items do not change structurally between time points; rather, the strength of these relations vary over time. In terms of unfavorable aspect-based factors, price dissatisfaction (item (1)), service dissatisfaction (item (12)), and meal dissatisfaction (item (8)) are all connected. Further, room dissatisfaction (item (4)), interior design dissatisfaction (item (10)), and bed dissatisfaction (item (14)) are connected. In addition, as shown by the connection strength, the main complaint is that the room design is so poor that it reflects negatively on reviews about the room and bed. Furthermore, the connection between room satisfaction (item (3)), interior design satisfaction (item (9)), and bed satisfaction (item (13)) is maintained over time. Therefore, we can conclude that the interior design influences the overall impression of the room and bed. Other factors are affected by the price such as dissatisfaction with the room and facilities. That is, there is constant connectivity between price dissatisfaction (item (1)), room dissatisfaction (item (4)), and facilities dissatisfaction (item (6)). As in the previous examples, Figure~\ref{hotelstats} shows that the observed summary statistics and simulated summary statistics are aligned (i.e., our model fits well).

\begin{figure}[htbp]
\begin{center}
\includegraphics[width = 1.0\textwidth]{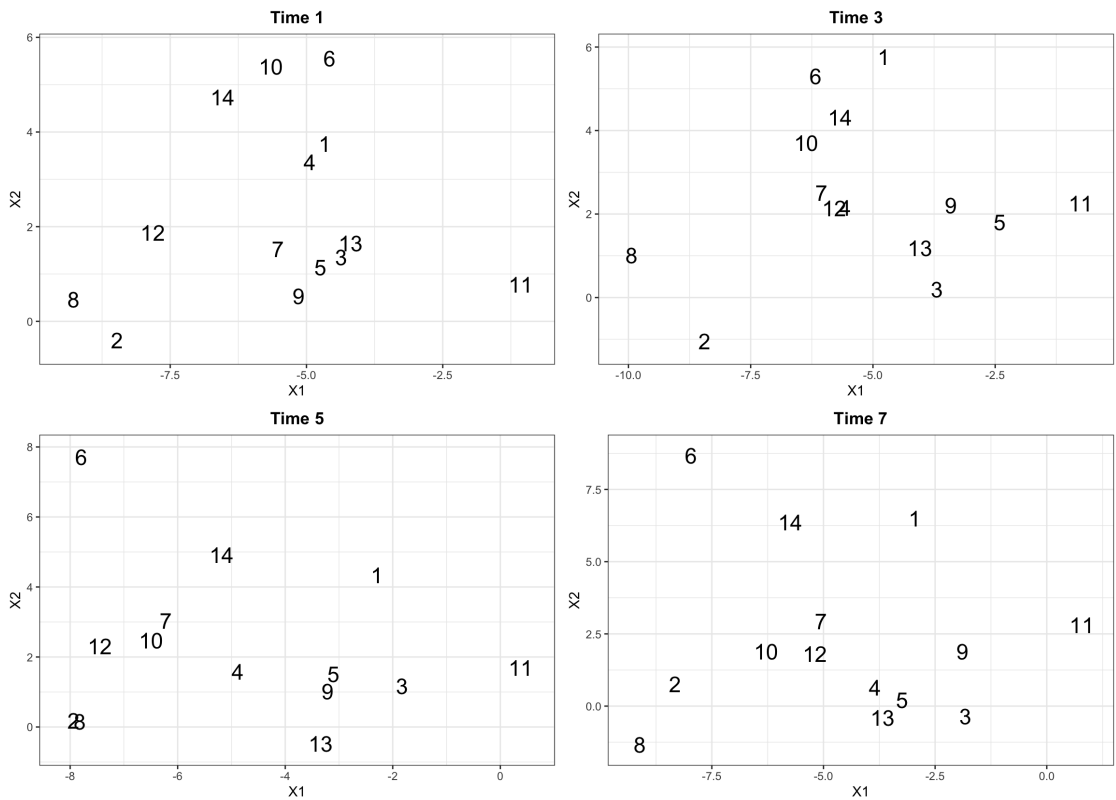}
\end{center}
\caption[]{Estimated latent positions for the hotel review data set. The closer two latent positions of items are, the more likely they have positive relations. }
\label{hoteldlsm-b}
\end{figure}

\paragraph{Comparison with DLSM-B} 
Figure \ref{hoteldlsm-b} shows the estimated latent positions from DLSM-B. Estimates are obtained from the posterior mean of 10,000 MCMC samples, which takes about 4 minutes. The latent position of item (3) becomes close to the latent positions of items (5), (9), (13) over time, which implies that they have positive temporal interactions. However, such interpretation is based on the relative distances between latent positions that can be subjective. Our method can be more informative in that FHS-DMH quantifies the change in interactions through $\gamma$ estimates. Table~\ref{hoteltable2} indicates that there are positive interactions between these items over time (i.e., $\gamma_{t,3,5}, \gamma_{t,3,9}, \gamma_{t,3,13}>0, \forall t$ ); FHS-DMH can determine statistically significant local interactions.

\section{Simulated Data Examples}

\begin{figure}[htbp]
\begin{center}
\includegraphics[width=1.0\textwidth]{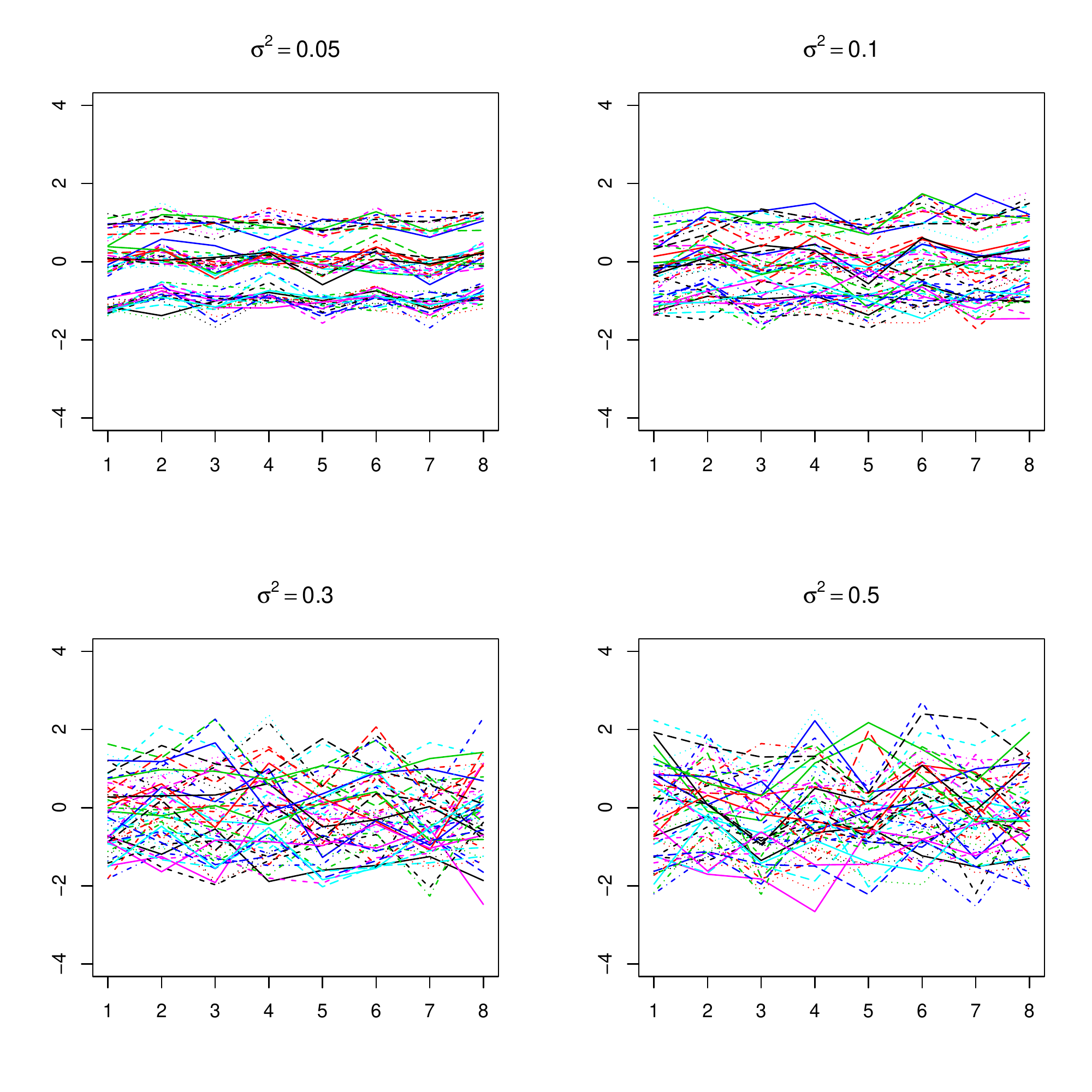}
\end{center}
\caption[]{Simulated functional parameters for FI-ERGMs under the four scenarios.}
\label{scenario}
\end{figure}

To validate our method, we conduct a simulation study under different scenarios. We simulate temporal binary response data sets with $n=600$ observations, $p=10$ items, and $t=8$ time points. This results in 55  functional parameters. We provide cyclic trends for the parameters using a function $\mu(t)=(1/8)\cos(\pi t) + (1/8)\sin(\pi t)$, which has fluctuating patterns around 0. We simulate 10 intercept parameters ($\alpha_{tj}$) from $N(-1+\mu(t), \sigma^2 \mathbf{I}_{8})$. Among the 45 interaction parameters ($\gamma_{tjk}$), 22 of them are simulated from $N(\mu(t), \sigma^2 \mathbf{I}_{8})$; these are assumed to be the true zero functions. Then, 11 of the interaction parameters are generated from $N(-1+\mu(t), \sigma^2 \mathbf{I}_{8})$ and the remaining 12 parameters are generated from $N(1+\mu(t), \sigma^2 \mathbf{I}_{8})$. Here, we consider four noise strength settings ($\sigma^2=0.05,0.1,0.3,0.5$). For the given model parameters $\lbrace \lbrace \alpha_{tj} \rbrace_{\forall t, j}, \lbrace \gamma_{tjk} \rbrace_{\forall t, j< k} \rbrace$, we simulate the data sets via $100n$ iterations of Metropolis-Hastings updates \citep{hunter2008ergm}. Figure~\ref{scenario} illustrates the simulated functional parameters under the different scenarios.

To study the performance of our method, we calculate the true positive rate (diagnose zero for the true zero functions) and true negative rate (diagnose nonzero for the true nonzero functions) for each scenario. Furthermore, we define the mean square error (MSE) as 
\[
\frac{1}{55}\sum_{i=1}^{55} (\widehat{\bm{\theta}}_{\bigcdot i} - \bm{\theta}_{\bigcdot i})'(\widehat{\bm{\theta}}_{\bigcdot i} - \bm{\theta}_{\bigcdot i}),
\]
where $\widehat{\bm{\theta}}_{\bigcdot i}$ is the functional parameter estimate obtained from FHS-DMH and $\bm{\theta}_{\bigcdot i}$ is the true parameter. Table~\ref{simultable} summarizes the results. Based on the MSE, we observe that FHS-DMH estimates are reasonably close to the true functional parameters. Furthermore, FHS-DMH can detect the true zero function well for reasonable noise settings ($\sigma^2=0.05, 0.1$). With increasing noise ($\sigma^2=0.3, 0.5$) in the simulated data, it becomes difficult to detect the true zero functions because the noises 
overwhelm the signals from each function.

\begin{table}[tt]
\centering
\begin{tabular}{cccc}
  \hline
Scenario  & MSE & TP & TN \\
  \hline
$\sigma^2 = 0.05$ & 0.09 & 1.00 & 0.92 \\
$\sigma^2 = 0.1$  & 0.12 & 0.86 & 0.91 \\
$\sigma^2 = 0.3$ & 0.16 & 0.68 & 0.76 \\
$\sigma^2 = 0.5$ & 0.23 & 0.73 & 0.67\\
\hline
\end{tabular}
\caption{10,000 MCMC samples are generated for each scenario; Monte Carlo standard errors are at 0.05. The MSE represents the mean square errors obtained from FHS-DMH estimates. TP (true positive) is the proportion of times that the true zero functional parameters are diagnosed as zero. TN (true negative) is the proportion of times that the true nonzero functional parameters are diagnosed as nonzero.}
\label{simultable} 
\end{table}

In addition, we study our method for different lengths of inner sampler (Step~2 in Algorithm~1). With the increasing length of the inner sampler, auxiliary variable samples become close to the stationary distribution at the expense of computational costs \citep{caimo2012bergm, park2018bayesian}. Under the same simulation setting above ($\sigma^2=0.05$), we calculate MSE, true positive rate, and true negative rate with the increasing length of the inner sampler. Table~\ref{auxiliarytest} indicates that the performances do not change much from $2n$. Therefore, we recommend using $2n$ as a practical choice for sampling auxiliary variables for FHS-DMH.

\begin{table}[tt]
\centering
\begin{tabular}{crrrr} \hline
Inner sampler & MSE & TP & TN  & Time(min)\\ \hline
$n$ & 0.12 & 1.00 & 0.85 & 21.01\\
$2n$ & 0.08 & 1.00 & 0.97 & 24.86 \\
$4n$ & 0.08 & 0.95 & 1.00 & 38.14 \\
$8n$ & 0.08 & 0.97 & 0.95 & 77.58\\
 \hline
\end{tabular}
\caption{MSE (mean square errors), TP (true positive), and TN (true negative) from the different lengths of the inner sampler in FHS-DMH.}
\label{auxiliarytest} 
\end{table}

\section{Discussion}

In this manuscript, we embed functional parameters in inhomogeneous exponential random graph models to study the temporal interactions among the items. Our models include intractable normalizing constants, and the number of functional parameters increases with an increasing number of items. We combine a double Metropolis-Hastings algorithm and a functional shrinkage method to address these computational and inferential challenges. Our study to real and simulated data examples shows that FHS-DMH can rule out weak interactions among items as well as provide a direct interpretation of temporal trends; our method can recover the dependence structure of the longitudinal networks well. To our knowledge, this is the first attempt to use functional shrinkage for models with intractable normalizing constants, which is an important contribution. 

Most ERGM-based dynamic models (e.g., TERGM, STERGM) describe probabilistic properties of the dynamic network by assuming Markov dependence between consecutive network observations with time-invariant model parameters. From different perspectives, we directly model the dynamic trends in the functional parameters using the nonparametric basis expansion $\boldsymbol{\Phi \beta}_{i}$ in \eqref{prior}. Our model assumes that the temporal dependence can be fully explained by time-varying parameter, and the networks on different time points are independent given the time-dependent parameter. This kind of  assumption is not uncommon; for instance, dynamic spatio-temporal models \citep[cf.][]{wikle2019spatio} also assume that observed processes are independent for a given time-evolving process which have similarities to our models. The practical advantage of our model is applicability to dynamic networks in continuous time as in \cite{lee2020varying}; it would be unclear to specify the Markov dependence when the time points are irregularly observed.


Similar to the variable selection methods \citep{nature14, park2019bayesian} for static networks, our methods are suited to data sets with a sufficient number of respondents ($n$) and a moderate number of items ($p$), as examples illustrated in this manuscript. Otherwise, it will suffer from a small $n$, large $p$ issues, which can lead to an unreliable inference. As a simple heuristic, we recommend applying our methods to problems with $n > p(p-1)/2$.

FHS-DMH is practical for moderate size of longitudinal item response data sets (e.g., $n=2,000$, $p=20$, $t=10$). With an increasing number of items and respondents, FHS-DMH becomes computationally expensive due to the auxiliary variable simulation in Step 2. Note that it is quite costly to collect large (for both $n,p$) item response data over a long period. It is usually the case that when the number of respondents increases, the number of items stays small, and when the number of items increases, the number of respondents stays small. Considering that it is challenging to have longitudinal item response data sets larger than our examples, FHS-DMH is applicable to many realistic cases. There have been several recent proposals to speed up inference for large network models. For instance, \cite{bouranis2017efficient} corrects MCMC samples collected from pseudo-posterior distribution, which is computationally efficient. \cite{park2019function} uses fast Gaussian process approximation to replace expensive Monte Carlo estimates for intractable normalizing constants. Adapting some of these methods would potentially lend itself to faster algorithms.

The computational methods developed here allow researchers in many disciplines to study temporal interactions among items for binary response data sets. Our methods could be applicable to a variable selection in existing temporal network models as well as for a broader class of functional models. Examples include temporal exponential random graph models \citep{hanneke2010discrete} and their variants \citep{krivitsky2014separable}, and functional regression models \citep{ramsay2007applied}.

\section*{Acknowledgement}
Jaewoo Park was partially supported by the Yonsei University Research Fund of 2020-22-0501 and the National Research Foundation of Korea (NRF-2020R1C1C1A0100386812). Ick Hoon Jin was partially supported by the Yonsei University Research Fund of 2019-22-0210 and the National Research Foundation of Korea (NRF-2020R1A2C1A01009881). Minsuk Shin was supported by the National Science Foundation through NSF-DMS-Statistics-2015528. The authors are grateful to Riccardo Rastelli for providing useful sample code. The authors are grateful to anonymous reviewers for their careful reading and valuable comments.

\bibliography{Reference}
\end{document}